\documentclass[sigconf]{acmart}
\usepackage{booktabs} % For formal tables
\usepackage{changes}
\usepackage[]{xcolor}
\definechangesauthor[name={Luca}, color=red]{luca}
\definechangesauthor[name={PaoloC}, color=red]{paoloc}
\definechangesauthor[name={PF}, color=red]{pf}
\definechangesauthor[name={dino}, color=blue]{dino}

\usepackage{listings}
\usepackage[english]{babel}
\usepackage{amsmath}
\usepackage{graphicx}
\usepackage{fancyhdr}
\usepackage{color}
\usepackage{caption}
\usepackage{multirow}

\colorlet{punct}{red!60!black}
\definecolor{background}{HTML}{EEEEEE}
\definecolor{delim}{RGB}{20,105,176}
\colorlet{numb}{black}

\lstdefinelanguage{json}{
    basicstyle=\normalfont\ttfamily,
    numbers=left,
    numberstyle=\scriptsize,
    stepnumber=1,
    numbersep=8pt,
    showstringspaces=false,
    breaklines=true,
    frame=lines,
    backgroundcolor=\color{background},
    literate=
     *{0}{{{\color{numb}0}}}{1}
      {1}{{{\color{numb}1}}}{1}
      {2}{{{\color{numb}2}}}{1}
      {3}{{{\color{numb}3}}}{1}
      {4}{{{\color{numb}4}}}{1}
      {5}{{{\color{numb}5}}}{1}
      {6}{{{\color{numb}6}}}{1}
      {7}{{{\color{numb}7}}}{1}
      {8}{{{\color{numb}8}}}{1}
      {9}{{{\color{numb}9}}}{1}
      {:}{{{\color{punct}{:}}}}{1}
      {,}{{{\color{punct}{,}}}}{1}
      {\{}{{{\color{delim}{\{}}}}{1}
      {\}}{{{\color{delim}{\}}}}}{1}
      {[}{{{\color{delim}{[}}}}{1}
      {]}{{{\color{delim}{]}}}}{1},
}

\setcopyright{rightsretained}
\acmDOI{10.475/123_4}

\acmISBN{123-4567-24-567/08/06}

\acmConference[]{}{}{}
\acmYear{}
\copyrightyear{}

\begin{document}
\title{PlayeRank: data-driven performance evaluation and player ranking in soccer via a machine learning approach}

\author{Luca Pappalardo}
\authornote{corresponding author}
\orcid{0000-0002-1547-6007}
\affiliation{%
  \institution{Institute of Information Science and Techologies (ISTI), CNR}
  \streetaddress{Via G. Moruzzi, 1}
  \city{Pisa}
  \state{Italy}
  \postcode{56124}
}
\email{luca.pappalardo@isti.cnr.it}

\author{Paolo Cintia}
\authornote{corresponding author}
\affiliation{%
  \institution{Department of Computer Science, University of Pisa}
  \streetaddress{Largo B. Pontecorso 3}
  \city{Pisa}
  \state{Italy}
  \postcode{56127}
}
\email{paolo.cintia@isti.cnr.it}

\author{Paolo Ferragina}
\affiliation{%
  \institution{Department of Computer Science, University of Pisa}
  \streetaddress{Largo B. Pontecorvo 3}
  \city{Pisa}
  \state{Italy}
  \postcode{56127}}
\email{ferragin@di.unipi.it}

\author{Emanuele Massucco}
\affiliation{%
  \institution{Wyscout}
  \city{Chiavari}
  \country{Italy}
}
\email{emanuele.massucco@wyscout.com}

\author{Dino Pedreschi}
\affiliation{%
 \institution{Department of Computer Science, University of Pisa}
  \streetaddress{Largo B. Pontecorso 3}
  \city{Pisa}
  \state{Italy}
  \postcode{56127}}
\email{pedre@di.unipi.it}

\author{Fosca Giannotti}
\affiliation{%
  \institution{Institute of Information Science and Techologies (ISTI), CNR}
  \streetaddress{Via G. Moruzzi, 1}
  \city{Pisa}
  \state{Italy}
  \postcode{56124}
}
\email{fosca.giannotti@isti.cnr.it}

% The default list of authors is too long for headers.
\renewcommand{\shortauthors}{L. Pappalardo et al.}

\begin{abstract}
The problem of evaluating the performance of soccer players is attracting the interest of many companies and the scientific community, thanks to the availability of massive data capturing all the events generated during a match (e.g., tackles, passes, shots, etc.). Unfortunately, there is no consolidated and widely accepted metric for measuring performance quality in all of its facets. 
In this paper, we design and implement {\sf PlayeRank}, a data-driven framework that offers a principled multi-dimensional and role-aware evaluation of the performance of soccer players. We build our framework by deploying a massive dataset of soccer-logs and consisting of millions of match events pertaining to four seasons of 18 prominent soccer competitions. By comparing {\sf PlayeRank} to known algorithms for performance evaluation in soccer, and by exploiting a dataset of players' evaluations made by professional soccer scouts, we show that {\sf PlayeRank} significantly outperforms the competitors.
We also explore the ratings produced by  {\sf PlayeRank} and discover interesting patterns about the nature of excellent performances and what distinguishes the top players from the others. At the end, we explore some applications of {\sf PlayeRank} --- i.e. searching players and player versatility --- showing its flexibility and efficiency, which makes it worth to be used in the design of a scalable platform for soccer analytics.

\end{abstract}

%
% The code below should be generated by the tool at
% http://dl.acm.org/ccs.cfm
% Please copy and paste the code instead of the example below.
%
 \begin{CCSXML}
<ccs2012>
<concept>
<concept_id>10002951.10003227.10003351</concept_id>
<concept_desc>Information systems~Data mining</concept_desc>
<concept_significance>500</concept_significance>
</concept>
</ccs2012>
\end{CCSXML}

\ccsdesc[500]{Information systems~Data mining}
\ccsdesc[500]{Information systems~Information retrieval}

\keywords{Sports Analytics, Clustering, Searching, Ranking, Multi-dimensional analysis, Predictive Modelling, Data Science, Big Data}

\maketitle

\section{Introduction}

Rankings of soccer players and data-driven evaluations of their performance are becoming more and more central in the soccer industry \cite{stein2017how,gudmundsson2017spatio,rein2016bigdata, bornn2018soccer}. 
On the one hand, many sports companies, websites and television broadcasters, such as Opta, WhoScored.com and Sky, as well as the plethora of online platforms for fantasy football and e-sports, widely use soccer statistics to compare the performance of professional players, with the purpose of increasing fan engagement via critical analyses, insights and scoring patterns. 
On the other hand, coaches and team managers are interested in analytic tools to support tactical analysis and monitor the quality of their players during individual matches or entire seasons. 
Not least, soccer scouts are continuously looking for data-driven tools to improve the retrieval of talented players with desired characteristics, based on evaluation criteria that take into account the complexity and the multi-dimensional nature of soccer performance. 
While selecting talents on the entire space of soccer players is unfeasible for humans as it is too much time consuming, data-driven performance scores could help in selecting a small subset of the best players who meet specific constraints or show some pattern in their performance, thus allowing scouts and clubs to analyze a larger set of players thus saving considerable time and economic resources, while broadening scouting operations and career opportunities of talented players.

The problem of data-driven evaluation of player performance and ranking are gaining interest in the scientific community too, thanks to the availability of massive data streams generated by (semi-)automated sensing technologies, such as the so-called soccer-logs \cite{stein2017how,gudmundsson2017spatio,rein2016bigdata, bornn2018soccer}, which detail all the spatio-temporal events related to players during a match (e.g., tackles, passes, fouls, shots, dribbles, etc.). Ranking players means defining a relation of order between them with respect to {\em some measure} of their performance over a sequence of matches. In turn, measuring performance means computing a \emph{data-driven performance rating} which quantifies the quality of a player's performance in a specific match and then aggregate them over the sequence of input matches. This is a complex task since there is no objective and shared definition of performance quality, which is an inherently multidimensional concept \cite{pappalardo2017human}.
Several data-driven ranking and evaluation algorithms have been proposed in the literature to date, but they suffer from three main limitations.

First, existing approaches are \emph{mono-dimensional}, in the sense that they propose metrics that evaluate the player's performance by focusing on one single aspect (mostly, passes or shots \cite{power2017passes, brooks2016developing,duch2010quantifying,pena2012network,lucey2014quality}), thus missing to exploit the richness of attached meta-information provided by soccer-logs. Conversely, soccer scouts search for a talented player based on "metrics" which combine many relevant aspects of their performance, from defensive skills to possession and attacking skills. Since mono-dimensional approaches cannot meet this requirement, there is the need for a framework capable to exploit a comprehensive evaluation of performance based on the richness of the meta-information available in soccer-logs.

Second, existing approaches evaluate performance without taking into account the specificity of each player's {\em role} on the field (e.g., right back, left wing), so they compare players that comply with different tasks \cite{duch2010quantifying,brooks2016developing,pena2012network,power2017passes,lucey2014quality}. Since it is meaningless to compare players which comply with different tasks and considering that a player can change role from match to match and even within the same match, there is the need for an automatic framework capable of assigning a role to players based on their positions during a match or a fraction of it.

Third, missing a gold standard dataset, existing approaches in the literature report judgments that consist mainly of informal interpretations based on some simplistic metrics (e.g., market value or goals scored \cite{torgler2007shapes,stanojevic2016towards,brooks2016developing}). It is important instead to evaluate the goodness of ranking and performance evaluation algorithms in a quantitative and throughout manner, through datasets built with the help of human experts as done for example for the evaluation of recommender systems in Information Retrieval. 

This paper presents the results of a joint research among academic computer scientists and data scientists of Wyscout \cite{wyscout}, the leading company for soccer scouting. 
The goal has been to study the limitations of existing approaches and develop {\sf PlayeRank}, a new-generation data-driven framework for the performance evaluation and the ranking of players in soccer. 
{\sf PlayeRank} offers a principled multi-dimensional and role-aware evaluation of the performance of soccer players, driven only by the massive and  standardized soccer-logs currently produced by several sports analytics companies (i.e., Wyscout, Opta, Stats). {\sf PlayeRank} is designed around the orchestration of the solutions to three main phases: a learning phase, a rating phase and a final ranking phase. {\sf PlayeRank} models the \emph{performance of a soccer player in a match} as a multidimensional vector of features extracted from soccer-logs. In the learning phase, {\sf PlayeRank} performs two main sub-tasks: (i) the \emph{extraction of feature weights}: since we do not have a ground-truth for ``learning'' the mapping from the performance features to the players' performance quality, we turn this problem into a classification problem between the multidimensional vector of features, aggregated over all players' of a {\em team}, and the result this team achieved in a match; (ii) the \emph{training of a role detector}: given that there are different player roles in soccer we identify, in an unsupervised way, a set of roles from the match events available in the soccer-logs. 

In the subsequent rating phase, the performance quality of a player in a match is evaluated as the scalar product between the previously computed feature weights and the values these feature get in that match played by that player. 
In the final ranking phase, {\sf PlayeRank} computes a set of \emph{role-based rankings} for the available players, by taking into account their performance ratings and their role(s) as they were computed in the two phases before. 

In order to validate our framework, we instantiated it over a massive dataset of soccer-logs provided by Wyscout which is unique in the large number of logged matches and players, and for the length of the period of observation. In fact, it includes 31 millions of events covering around 20K matches and 21K players in the last four seasons of 18 prominent soccer competitions: La Liga (Spain), Premier League (England), Serie A (Italy), Bundesliga (Germany), Ligue 1 (France), Primeira Liga (Portugal), Super Lig (Turkey), Souroti Super League (Greece), Austrian Bundesliga (Austria), Raiffeisen Super League (Switzerland), Russian Football Championship (Russia), Eredivisie (The Netherlands), Superliga (Argentina), Campeonato Brasileiro S\'erie A (Brazil), UEFA Champions League, UEFA Europa League, FIFA World Cup 2018 and UEFA Euro Cup 2016. Then we performed an extensive experimental analysis advised by a group of professional soccer scouts which showed that {\sf PlayeRank} is robust in agreeing with a ranking of players given by these experts, with an improvement up to 30\% (relative) and 21\% (absolute) with respect to the current state-of-the-art algorithms \cite{duch2010quantifying,brooks2016developing}. 

One of the main characteristics of {\sf PlayeRank} is that, by providing a score which meaningfully synthesizes a player's performance quality in a match or in a series of matches, it enables the analysis of the statistical properties of player performance in soccer. In this regard, the analysis of the performance ratings resulting from {\sf PlayeRank}, for all the players and all the matches in our dataset, revealed several interesting patterns.

First, on the basis of the players' average position during a match, the role detector finds \emph{eight} main roles in soccer (Section \ref{sec:role_detector}) and enables the investigation of the notion of {\em player's versatility}, defined as his ability to change role from match to match (Section \ref{sec:versatility}). Second, the analysis of feature weights reveals that there is no significant difference among the 18 competitions, with the only exception of  the competitions played by national teams (Section \ref{sec:feature_weighting}). Third, the distribution of player ratings changes by role, thus suggesting that the performance of a player in a match highly depends on the zone of the soccer field he is assigned to  (Section \ref{sec:ratings}). This is an important aspect that will be exploited to design a novel search engine for soccer players (Section \ref{sec:retrieving_results}). Fourth, we find that the distribution of performance ratings is strongly peaked around its average, indicating that ``outlier'' performances are rare (Section \ref{sec:ratings}). In particular, these outlier performances are unevenly distributed across the players: while the majority of players achieve a few excellent performances, a tiny fraction of players achieve many excellent performances. Moreover, we find that top players do not always play in an excellent way but, nonetheless, they achieve excellent performances more frequently than the other players (Section \ref{sec:ratings}).

In conclusion, our study and experiments show that {\sf PlayeRank} is an innovative data-driven and open-source framework which goes beyond the state-of-the-art results in the evaluation and ranking of soccer players.\footnote{The source code of {\sf PlayeRank} and a portion of the soccer-logs used to train it will be made available in the camera-ready version of this paper.} This study also provides the first thorough, and somewhat surprising, characterization of soccer performance. The last section will start from {\sf PlayeRank} and its study to present a set of new challenging problems in soccer analytics that we state and comment in order to stimulate the research interest from the community of data scientists.

\section{Related Works}
\label{sec:related}

The availability of massive data portraying soccer performance has facilitated recent advances in soccer analytics. The so-called soccer-logs \cite{stein2017how,gudmundsson2017spatio,rein2016bigdata, bornn2018soccer}, capturing all the events occurring during a match, are one of the most common data format and have been used to analyze many aspects of soccer, both at team \cite{lucey2013assessing,cintia2015harsh,pappalardo2017quantifying,wang2015discerning, decroos2018automatic} and individual level \cite{brooks2016developing,duch2010quantifying,nsolo2018player}. Among all the open problems in soccer analytics, the data-driven evaluation of a player's performance quality is the most challenging one, given the absence of a ground-truth for that performance evaluation. 

\paragraph{Data-driven evaluation of performance.} While many metrics have been proposed to capture specific aspects of soccer performance (e.g., expected goals, pass accuracy, etc.), just a few approaches evaluate a player's performance quality in a systemic way.

The flow centrality (FC) metric proposed by Duch et al.\ \cite{duch2010quantifying}, one of the first attempts in this setting, is defined as the fraction of times a player intervenes in pass chains which end in a shot. Based on this metric, they rank all players in UEFA European Championship 2008 and observe that 8 players in their top-20 list belong to the UEFA's top-20 list which was released just after the competition. Being based merely on pass centrality, as the authors themselves highlight in the paper, the FC metric mostly makes sense for midfielders and forwards. 

Brooks et al.\ \cite{brooks2016developing} develop the Pass Shot Value (PSV), a metric to estimate the importance of a pass for generating a shot. They represent a pass as a vector of 360 features describing the vicinity of a field zone to the pass' origin and destination. Then, they use a supervised machine-learning model to predict whether or not a given pass results in a shot. The feature weights resulting from the model training are used to compute PSV as the sum of the feature weights associated with the pass' origin and destination. They finally used soccer-logs to rank players in La Liga 2012-13 according to their average PSV, showing that it correlates with the rankings based on assists and goals. Unfortunately, as the authors highlight in the paper, PSV is strongly biased towards offensive-oriented players. Moreover, PSV is a pass-based metric which thus omits all the other kinds of events observed during a soccer match, and lacks of a proper validation.

Instead of proposing their own algorithm for performance quality evaluation, Nsolo et al. \cite{nsolo2018player} extract performance metrics from soccer-logs to predict the WhoScored.com performance rating with a machine learning approach. The resulting model is more accurate for specific roles (e.g., forwards) and competitions (e.g., English Premier League) when predicting if a player is in the top 10\%, 25\% and 50\% of the WhoScored.com ranking. 

The problem of evaluating players performance got much attention in other team sports, like hockey, basketball and especially baseball. In hockey, Schulte and Zhao proposed the Scoring Impact metric (SI) \cite{schulte2017apples} to rank ice hockey players in NHL depending on his team's chance of scoring the next goal. In basketball, the Performance Efficiency Rating\footnote{https://www.basketball-reference.com/about/per.html} is nowadays a widely used metric to assess players' performance by deploying basketball-logs (i.e. pass completed, shots achieved, etc.). In baseball, a plethora of statistical metrics have been proposed to evaluate the performance of  players and teams \cite{baumer2014sabermetric}.

\paragraph{Rating systems for sports teams.}
Many studies also focused on developing the so-called \emph{rating systems}, like Elo and TrueSkill \cite{lasek2013predictive,herbrich2006trueskill}, which rank teams or players based on their past victories/defeats and the estimated strength of the opponent. Therefore, they do not take into account neither player-observed match events nor other quantitative aspects of individual and collective performance \cite{pappalardo2017quantifying}. As a result, unlike {\sf PlayeRank}, such rating systems are unable to provide an explicit characterization of the evaluated performance of a player as well as to discern his contribution in a match. 

\paragraph{Relations between performance and market value.}
Another strand of literature focuses on quantifying the relation between proxies of a player's quality, like market value, wage or popularity, and his performance on the field. Stanojevic and Gyarmati \cite{stanojevic2016towards} use soccer-logs to infer the relation between a player's typical performance and his market value as estimated by crowds. They find a large discrepancy between estimated and real market values, due to the lack of important information such as injury-proneness and commercialization capacity. M\"uller et al.\ \cite{muller2017beyond} develop a similar approach and use soccer-logs, as well as players' popularity data and market values in the preceding years, to estimate a player transfer fee. They show that for the low- and medium-priced players the estimated market values are comparable to estimations by the crowd, while the latter performs better for the high-priced players. Torgler and Schmidt \cite{torgler2007shapes} investigate what shapes performance in soccer, represented as a player number of goals and assists. They find that salary, age and team effects have a statistically significant impact on a player performance on the field. 

\paragraph{Position of our work}
Despite an increasing interest in this research field, our review of the state-of-the-art highlights that there is no validated framework allowing for a multi-dimensional and role-aware evaluation of soccer performance quality. In this paper, we overcome this issue by proposing {\sf PlayeRank}, a framework that deploys all the possible events described by soccer-logs to evaluate player's performance quality and player's role in a match. In contrast to FC and PSV, which lack of a proper validation with domain experts, we test the framework against a humanly-labeled dataset we have specifically built for the purpose of evaluating soccer players performance. Finally, and for the first time in the literature, we shed some light on the statistical patterns that characterize soccer players performance by providing a novel and throughout analysis that exploits {\sf PlayeRank} scores and the large and unique dataset of competitions, teams and players Wyscout made available to us. 

\section{The PlayeRank framework}
\label{sec:framework}

\begin{figure}
\centering
\includegraphics[scale=0.34]{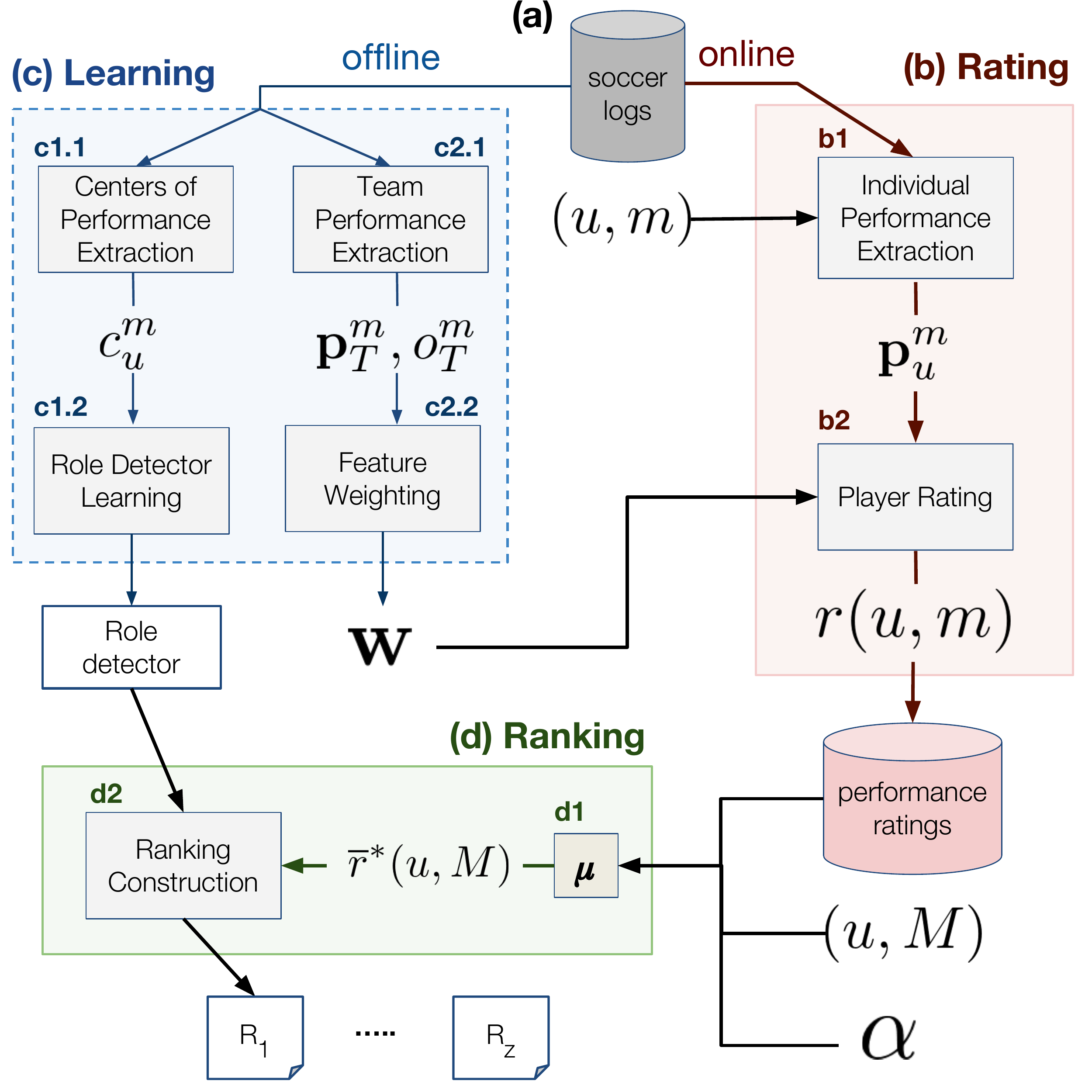}
\caption{Schema of the {\sf PlayeRank} framework. Starting from a database of soccer-logs (a), it consists of three main phases. The learning phase (c) is an "offline" procedure: it must be executed at least one before the other phases since it generates information used in the other two phases, but then it can be updated separately. The rating (b) and the ranking phases (d) are online procedures, i.e., they are executed every time a new match is available in the database of soccer-logs. We refer to the text for the notation used in the figure.}
\label{fig:framework_schema}
\end{figure}

Figure \ref{fig:framework_schema} describes how the {\sf PlayeRank} framework operates. It is designed to work with soccer-logs, in which a soccer match consists of a sequence of events encoded as a tuple: $\langle id, type, position, timestamp\rangle$, where $id$ is the identifier of the player which originated/refers to this event, $type$ is the event type (i.e., passes, shots, goals, tackles, etc.), $position$ and $timestamp$ denote the spatio-temporal coordinates of the event over the soccer field. {\sf PlayeRank} assumes that soccer-logs are stored into a database, which is updated with new events after each soccer match (Figure \ref{fig:framework_schema}a).

The key task addressed by {\sf PlayeRank} is the \emph{``evaluation of the performance quality of a player $u$ in a soccer match $m$''}. This consists of computing a numerical rating $r(u, m)$, called \emph{performance rating}, that aims at capturing the quality of the performance of $u$ in $m$ given {\em only} the set of events related to that player in that match. This is a complex task because of the many events observed in a match, the interactions among players within the same team or against players of the opponent team, and the fact that players performance is inextricably bound to the performance of their team and possibly of the opponent team. {\sf PlayeRank} addresses such complexity by means of a procedure which hinges onto a massive database of soccer-logs and consists of three phases: a rating phase, a ranking phase and a learning phase.

\subsection{Rating phase}
\label{sec:rating_phase}
The rating phase (step b in Figure \ref{fig:framework_schema}) is the procedure responsible for the computation of the performance rating $r(u, m)$ and it is run for each player $u$ every time a new match $m$ becomes available in the soccer-logs database. This phase exploits information computed "offline" and consists of two main steps: individual performance extraction (Figure \ref{fig:framework_schema}, step b1) and player rating (Figure \ref{fig:framework_schema}, step b2).

\subsubsection*{\bf Individual Performance Extraction}
Given that a match $m$ is represented as a set of events, {\sf PlayeRank} {\em models} the performance of a player $u$ in $m$ by means of a $n$-dimensional feature vector $\mathbf{p}_u^m = [x_1, \dots, x_{n}]$, where $x_i$ is a feature that describes a specific aspect of $u$'s behavior in match $m$ and is computed from the set of events played by $u$ in that match. 
In our experiments at Section \ref{sec:experiments}, we provide an example of $n{=}76$ features extracted from the Wyscout dataset. Some features count some events (e.g., number of fouls, number of passes, etc.), some others are at a finer level in that they distinguish the outcome of those events --- i.e., if they were ``accurate'' or ``not accurate''. 
Note that {\sf PlayeRank} is designed to work with any set of features, thus giving to the user a high flexibility about the description and deployment of soccer performance.

\subsubsection*{\bf Player Rating}
\label{sec:rating}
The evaluation of the performance of a player $u$ \emph{in a single match} $m$ is computed as the scalar product between the values of the features referring to match $m$ and the feature weights $\mathbf{w}$ computed during the learning phase (Figure \ref{fig:framework_schema}, step c2.2, described in the next Section \ref{sec:learning}). Each feature weight models the importance of that feature in the evaluation of the performance quality of any player. 

Formally speaking, given the multi-dimensional vector of features $\mathbf{p}_u^m = [x_1, \dots, x_n]$ and their weights $\mathbf{w}$, {\sf PlayeRank} evaluates the performance of a player $u$ in a match $m$ as follows:
\begin{equation}
r(u, m) = \frac{1}{R}\; \sum_{i = 1}^n w_i \,\times\,  x_i.
\end{equation}

The quantity $r(u, m)$ is called the \emph{performance rating} of $u$ in match $m$, where $R$ is a normalization constant such that $r(u, m) \in [0,1]$.
Since we decided to not include the number of goals scored in a match into the set of features, for reasons that are explained in Section \ref{sec:learning} (learning phase), but goals could themselves be important to evaluate the performance of some (offensive) players, {\sf PlayeRank} can be adapted to manage goals too via an \emph{adjusted-performance rating}, defined as follows: 
\begin{equation}
r^*(u, m) = \alpha \;\times\; norm\_goals + (1 - \alpha) \;\times\; r(u, m)
\label{eq:r_star}
\end{equation}
where $norm\_goals$ indicates the number of goals scored by $u$ in match $m$ normalized in the range $[0, 1]$, and $\alpha {\in} [0, 1]$ is a parameter indicating the importance given to goals into the new rating. Clearly, $r^*(u, m) {=} r(u, m)$ when $\alpha {=} 0$, and $r^*(u, m) {=} norm\_goals$ when $\alpha {=} 1$.

\smallskip Finally, {\sf PlayeRank} computes the rating of a \emph{player} $u$ over a series of matches $M = (m_1, \dots, m_g)$ by aggregating $u$'s ratings over those matches according to a function $\mu(r(u, m_1), \dots, r(u, m_g))$ which, in this paper, is set to the Exponential Weighted Smoothing Average (EWMA). This way, the performance quality of player $u$ after $g$ matches is computed as: 
\begin{equation}
\overline{r}(u, M) = \overline{r}(u, m_g) = \beta \times r(u, m_g) + (1 - \beta) \times  \overline{r}(u, m_{g - 1})
\label{eq:r_mean}
\end{equation}
where $\beta$ is a proper smoothing factor set in the range $[0,1]$. In other words, the performance quality of player $u$ after $g$ matches, i.e. $\overline{r}(u, m_g)$, is computed as the weighted average of the rating $r(u, m_g)$ reported by $u$ in the last match $m_g$ and the previous smoothed ratings $\overline{r}(u, m_{g - 1})$. This way we are counting more the recent performances of players. Similarly, the \emph{goal-adjusted rating} $\overline{r}^*(u, m_g)$ of $u$ given a series of $g$ matches is computed as the EWMA of his adjusted performance ratings. The quantity $\overline{r}(u, M)$ is called the \emph{player rating} of player $u$ given $M$, while $\overline{r}^*(u, M)$ is called the \emph{adjusted-player rating} of player $u$ given $M$.

\subsection{Ranking phase.}
Based on the {\em players ratings} computed in the previous phase, {\sf PlayeRank} constructs a set of \emph{role-based rankings} $R_1, \dots, R_z$, each corresponding to one of the $z$ roles identified by a \emph{role detector} (step c1.2, described in the next Section \ref{sec:learning}), an algorithm previously trained during the learning phase which assigns to one or more roles each player $u$ in a match $m$. 
{\sf PlayeRank} assigns a player $u$ to $R_i$ if he has at least $x$\% of the matches in $M$ assigned to role $i$, where $x$ is a parameter chosen by the user. 
In our experiments at Section \ref{sec:experiments} we select $x = 40\%$, a choice dictated by the fact that arguably a soccer player may be assigned to at most two roles. Experiments showed this threshold is robust, however this parameter can be chosen by the user when running {\sf PlayeRank}, possibly increasing the number of assigned roles per player (i.e., his versatility). Depending on the value of the threshold $x$, a player can appear in more than one ranking and with different ranks since they depend on $\overline{r}(u, M)$. 

\subsection{Learning phase.}
\label{sec:learning}
The learning phase (Figure \ref{fig:framework_schema}{c}) is executed "offline"in order to generate information used in the rating and the ranking phases. It
consists of two main steps: feature weighting and role detector training. 
\subsubsection*{\bf Feature weighting}
Performance evaluation is a difficult task because we do not have an objective evaluation of the performance $\mathbf{p}_u^m$ of each individual player $u$. This technically means that we do not have a ground-truth dataset to learn a {\em relation} between performance features and performance quality of $u$ in match $m$. 
On the other hand, we observe that the outcome of a match may be considered a natural proxy for evaluating performance quality at \emph{team} level. Therefore, we overcome that limitation by proposing a {\em supervised} approach: we determine the impact of the $n$ chosen features onto a player performance by looking in turn at the team-wise contribution of these features to the match outcome.

This idea is motivated by the fact that (i) a team's ultimate purpose in a match is to win by scoring one goal more than the opponent, (ii) some actions of players during a match have a higher impact on the chances of winning a match than others. For example,  making a pass which puts a teammate in condition to score a goal (assist) is intuitively more valuable than making a pass to a close teammate in the middle of the field. Conversely, getting a red card is intuitively less valuable than, let's say, winning a dribble against an opponent. Therefore those actions which strongly increase (or decrease) the chances of winning a match must be evaluated more during the evaluation, either positively or negatively. While soccer practitioners and fans have in mind an idea of what the most and the least valuable actions during a match are, it is important to develop a data-driven and automatic procedure that quantifies how much valuable an action is with respect to increasing or decreasing the chances of winning a match.

{\sf PlayeRank} implements this syllogism via a two-phase approach. In the first phase (Figure \ref{fig:framework_schema}, step c2.1) it extracts the performance vector $\mathbf{p}_T^m$ of team $T$ in match $m$ and the outcome $o_T^m$ of that match: where $o_T^m = 1$ indicates a victory for team $T$ in match $m$ and $o_T^m = 0$ indicates a non-victory (i.e., a defeat or a draw) for $T$. The team performance vector $\mathbf{p}_T^m = [x_1^{(T)}, \dots, x_n^{(T)}]$ is obtained by summing the corresponding features over all the players $U_T^m$ composing team $T$ in match $m$: 
$$ \mathbf{p}_T^m[i] = \sum_{u \in U_T^m} \mathbf{p}_u^m[i].$$

\noindent In the second phase (Figure \ref{fig:framework_schema}, step 2.2), {\sf PlayeRank} solves a classification problem between the team performance vector $\mathbf{p}_T^m$ and the outcome $o_T^m$. This classification problem has been shown in \cite{pappalardo2017quantifying} to be meaningful, because there is a strong relation between the team performance vector and the match outcome. 
We use a linear classifier, such as the Linear Support Vector Machine (SVM), to solve the previous classification problem and then we extract from the classifier the weights $\mathbf{w} = [w_1, \dots, w_n]$ which quantify the influence of the features to the outcomes of soccer matches, as explained above. 

These weights are then used in the rating phase (Figure \ref{fig:framework_schema}, step b2) to compute the performance ratings of players. 
%Note that the set of features at team level used in this step must be the same set of features at individual level used in step b2 of the rating phase, in order to guarantee consistency between the two phases.

\subsubsection*{\bf Role detector training}
As pointed out in \cite{schulte2017apples, pettigrew2015assessing}, performance ratings are meaningful only when comparing players with similar {\em roles}. In soccer, each role corresponds to a different area of the playing field where a player is assigned responsibility relative to his teammates \cite{bialkowski2014analysis}. Different roles imply different tasks, hence it is meaningless to compare, for example, a player that is asked to create goal occasions and a player that is asked to prevent the opponents to score. Furthermore, a role is not a unique label as a player's area of responsibility can change from one match to another and even within the same match.
Given these premises, we decided to design and implement an algorithm able to detect the role associated with a player's performance in a match based on the soccer-logs. We observe that they do exist methods, originally designed for hockey \cite{schulte2017apples}, that compute the roles of players via an affinity clustering applied over a heatmap describing their presence in predefined zones of the field. But these approaches are arguably not effective in soccer because it offers a lower density of match events w.r.t. hockey. Nonetheless we experimented and discarded the approach of \cite{schulte2017apples} because it produces on our dataset a clustering with a very low quality (i.e., silhouette score $ss < 0.2$). 

Conversely, {\sf PlayeRank} detects the role of a player $u$ in a match $m$ by looking at his {\em average position}. This is motivated by the fact that a player's role is often defined as the position covered by the player relative to his teammates \cite{bialkowski2014analysis}. 
This is called the {\em center of performance} for $u$ in $m$ and it is denoted as $\mathbf{c}_u^m = (\overline{x}_u^m, \overline{y}_u^m)$, where $\overline{x}_u^m$ and $\overline{y}_u^m$ are the average coordinates of $u$'s events in match $m$, as they are extracted from the soccer-logs (Figure \ref{fig:framework_schema}, step c1.1). 
Then {\sf PlayeRank} deploys a $k$-means algorithm \cite{hartigan1979kmeans} to group the centers of performance of all players $u$ in all matches $m$ (Figure \ref{fig:framework_schema}, step 1.2). 

{\sf PlayeRank} also accounts for the possibility of having ``hybrid'' roles where a center of performance is assigned to two or more clusters. 
This is useful in situations where the center of performance of a player $u$ is between two or more clusters, and so the role of $u$ in match $m$ cannot be well characterized by just one single cluster.
Therefore, {\sf PlayeRank} aims at a finer classification of roles via a {\em soft clustering}. For every center of performance $\mathbf{c}_u^m$ occurring in some cluster $C_i$, {\sf PlayeRank} computes its $k$-silhouette $s_k(\mathbf{c}_u^m)$ with respect to every other cluster $C_k$ ($k \neq i$) as: 
\begin{equation}
s_k(\mathbf{c}_u^m) = \frac{\overline{d_k}(\mathbf{c}_u^m) - \overline{d_i}(\mathbf{c}_u^m)}{max(\overline{d_i}(\mathbf{c}_u^m), \overline{d_k}(\mathbf{c}_u^m))},
\end{equation}
where $\overline{d_z}(\mathbf{c}_u^m)$ is the average distance between $\mathbf{c}_u^m$ and all other points in cluster $C_z$. {\sf PlayeRank} assigns $\mathbf{c}_u^m$ to {\em every} cluster $C_j$ for which $s_j(\mathbf{c}_u^m) \leq \delta_s$, where $\delta_s$ is a threshold indicating the tolerance to ``hybrid'' centers. If no such $j$ does exist, $\mathbf{c}_u^m$ is assigned to the cluster $C_i$ given by the partitioning computed by the $k$-means algorithm. 

For the sake of completeness we mention that in approaching the task of role classification we have considered other,  more sophisticated modeling of players' performance such as heatmaps (as in \cite{schulte2017apples}, see comments above) or events direction (as in \cite{bialkowski2014analysis}), but clusters were of lower quality in terms of the silhouette score. 

\section{Experimental Results}
We implemented the {\sf PlayeRank} framework and executed it on a massive database of soccer-logs provided by the company Wyscout \cite{wyscout}. In this section, we show experiments for each of the modules described in Section \ref{sec:framework} and depicted in Figure \ref{fig:framework_schema}.

\label{sec:experiments}
\subsection{Wyscout dataset}
We use a database of soccer-logs provided by Wyscout consisting of 31,496,332 events, capturing $19,619$ matches, 296 clubs and $21,361$ players of several seasons of 18 prominent competitions around the world (see Table \ref{tab:competitions}): La Liga (Spain), Premier League (England), Serie A (Italy), Bundesliga (Germany), Ligue 1 (France), Primeira Liga (Portugal), Super Lig (Turkey), Souroti Super League (Greece), Austrian Bundesliga (Austria), Raiffeisen Super League (Switzerland), Russian Football Championship (Russia), Eredivisie (The Netherlands), Superliga (Argentina), Campeonato Brasileiro S\'erie A (Brazil), UEFA Champions League, UEFA Europa League, FIFA World Cup 2018 and UEFA Euro Cup 2016. 

\begin{table*}[]
\begin{tabular}{lll||r|r|r|r|}
\hline
\multicolumn{1}{|l|}{\textbf{\Large competition}}          & \multicolumn{1}{l|}{\textbf{\Large area}} & \textbf{\Large type} & \textbf{\Large \#seasons} & \textbf{\Large \#matches} & \textbf{\Large \#events} & \textbf{\Large \#players} \\ \hline
\multicolumn{1}{|l|}{La Liga}                       & \multicolumn{1}{l|}{Spain}            & national      & 4                & 1520             & 2,541,873       & 1264             \\ \hline
\multicolumn{1}{|l|}{Premier League}                & \multicolumn{1}{l|}{England}          & national      & 4                & 1520             & 2,595,808       & 1231             \\ \hline
\multicolumn{1}{|l|}{Serie A}                       & \multicolumn{1}{l|}{Italy}            & national      & 4                & 1520             & 2,610,908       & 1499             \\ \hline
\multicolumn{1}{|l|}{Bundesliga}                    & \multicolumn{1}{l|}{Germany}          & national      & 4                & 1124             & 2,075,483       & 1042             \\ \hline
\multicolumn{1}{|l|}{Ligue 1}                       & \multicolumn{1}{l|}{France}           & national      & 4                & 1520             & 2,592,708       & 1288             \\ \hline
\multicolumn{1}{|l|}{Primeira Liga}                 & \multicolumn{1}{l|}{Portugal}         & national      & 4                & 1124             & 1,720,393       & 1227             \\ \hline
\multicolumn{1}{|l|}{Super Lig}                     & \multicolumn{1}{l|}{Turkey}           & national      & 4                & 1124             & 1,927,416       & 1182             \\ \hline
\multicolumn{1}{|l|}{Souroti Super Lig}             & \multicolumn{1}{l|}{Greece}           & national      & 4                & 1060             & 1,596,695       & 1151             \\ \hline
\multicolumn{1}{|l|}{Austrian Bundesliga}           & \multicolumn{1}{l|}{Austria}          & national      & 4                & 720              & 1,162,696       & 593              \\ \hline
\multicolumn{1}{|l|}{Raiffeisen Super League}       & \multicolumn{1}{l|}{Switzerland}      & national      & 4                & 720              & 1,124,630       & 647              \\ \hline
\multicolumn{1}{|l|}{Football Championship}         & \multicolumn{1}{l|}{Russia}           & national      & 4                & 960              & 1,593,703       & 1046             \\ \hline
\multicolumn{1}{|l|}{Eredivisie}                    & \multicolumn{1}{l|}{The Netherlands}  & national      & 4                & 1248             & 2,021,164       & 1177             \\ \hline
\multicolumn{1}{|l|}{Superliga}                     & \multicolumn{1}{l|}{Argentina}        & national      & 4                & 1538             & 2,450,170       & 1870             \\ \hline
\multicolumn{1}{|l|}{Campeonato Brasileiro Serie A} & \multicolumn{1}{l|}{Brazil}           & national      & 4                & 1437             & 2,326,690       & 1790             \\ \hline
\multicolumn{1}{|l|}{UEFA Champions League}         & \multicolumn{1}{l|}{Europe}           & continental   & 3                & 653              & 995,363         & 3577             \\ \hline
\multicolumn{1}{|l|}{UEFA Europa League}            & \multicolumn{1}{l|}{Europe}           & continental   & 3                & 1416             & 1,980,733       & 9100             \\ \hline
\multicolumn{1}{|l|}{UEFA Euro Cup 2016}            & \multicolumn{1}{l|}{Europe}           & continental   & 1                & 51               & 78,140          & 552              \\ \hline
\multicolumn{1}{|l|}{FIFA World Cup 2018}           & \multicolumn{1}{l|}{World}            & international & 1                & 64               & 101,759         & 736              \\ \hline \hline
                                                    &                                       &               & 64               & 19,619           & 31,496,332      & (*)21,361           \\ \cline{4-7} 
\end{tabular}
\caption{List of competitions with the corresponding geographic area, type and total number of seasons, matches, events and players. The dataset covers 18 competitions, for a total of 64 soccer seasons and around 20K matches, 31M events and 21K players. (*) 21,361 indicates the number of distinct players in the dataset, as some players play with their teams in both national and continental/international competitions.}
\label{tab:competitions}
\end{table*}

\begin{figure}[htb]\centering
\includegraphics[scale=0.38]{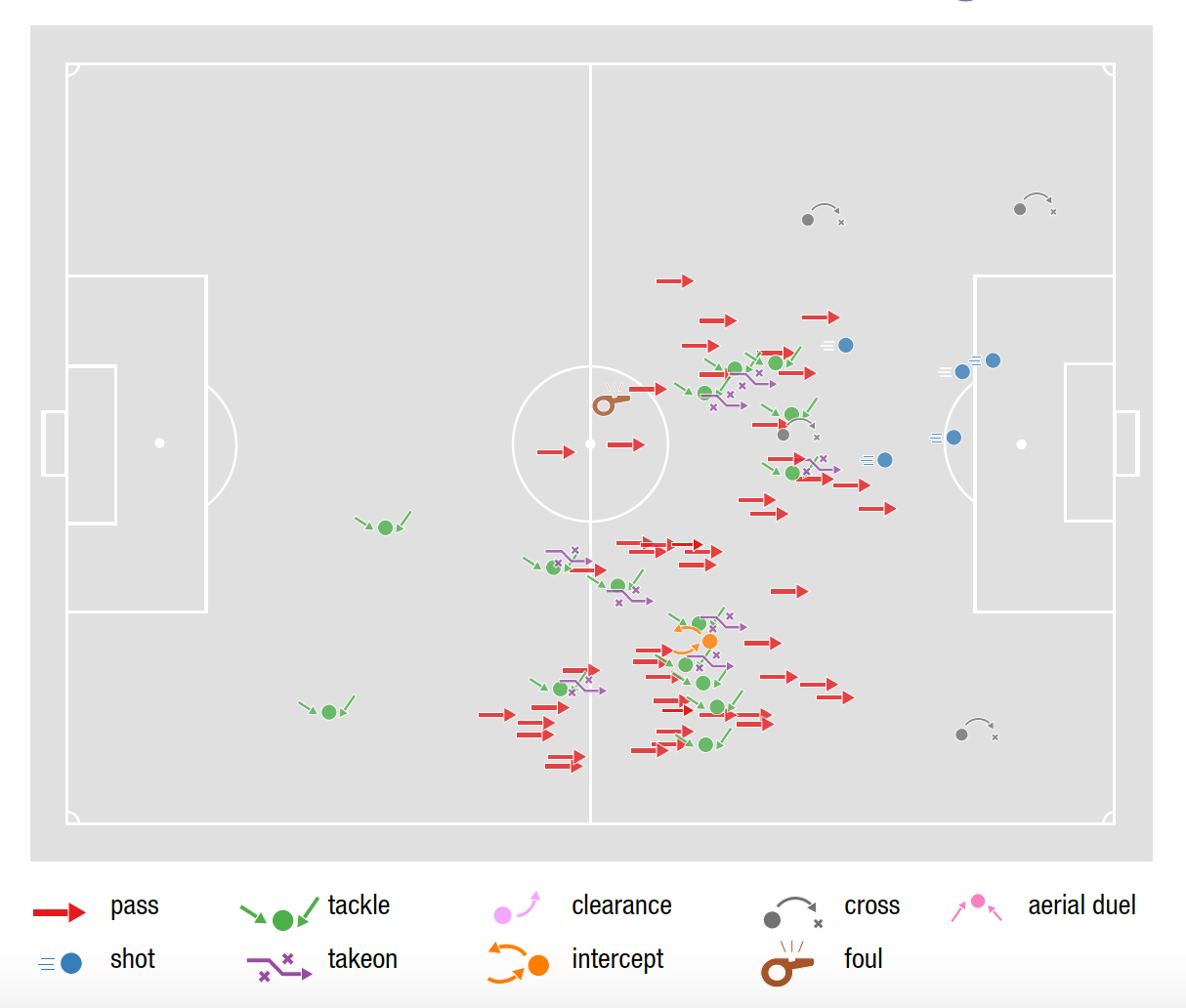}
\caption{Events observed for Lionel Messi (FC Barcelona) during a match in La Liga (Spain), season 2015/2016. Each event is shown on the field at the position where it has occurred with a marker indicating the type of the event.}
\label{fig:example_events}
\end{figure}

Each event records: (i) a unique event identifier; (ii) the type of the event; (iii) a time-stamp; (iv) the player related to the event; (v) the team of the player; (vi) the match in which the event is observed; (vii) the position on the soccer field, specified by a pair of integers in the range $[0, 100]$ indicating the percentage from the left corner of the attacking team; (viii) the event subtype and a list of tags, that enrich the event with additional information (see Table~\ref{tab:events}). We do not consider the {\em goalkeeping} events available from the Wyscout APIs, as we discard goalkeepers from the analysis.\footnote{Goalkeepers would need a dedicated analysis since it is the only role having different game rules w.r.t. to all other players.} Figure \ref{fig:example_event} shows an example of an event in the dataset, corresponding to an accurate pass by player 3344 (Rafinha) of team 3161 (Internazionale) made at second 2.41 in the first half of match 2576335 (Lazio - Internazionale) started at position (49, 50) of the field. Figure \ref{fig:example_events} shows a pictorial representation of the events produced by player Lionel Messi during a match in the Spanish La Liga, where each event is drawn at the position of the field where it has occurred.
  
\begin{figure}[htb]\centering
\begin{lstlisting}[language=json, numbers=none]
 {"id": 253668302,
 "eventName": "Pass",
 "eventSec": 2.41,
 "playerId": 3344,
 "matchId": 2576335,
 "teamId": 3161,
 "positions": [{"x": 49, "y": 50}],
 "subEventId": 85,
 "subEventName": "Simple pass",
 "tags": [{"id": 1801}]}
 \end{lstlisting}
\caption{Example of event in the dataset corresponding to an accurate pass by player 3344 (Rafinha) of team 3161 (Internazionale) made at second 2.41 of match 2576335 (Lazio - Internazionale) started at position (49, 50) of the field.}
\label{fig:example_event}
\end{figure}

\begin{table*}[htb]\centering
\def\arraystretch{1.5}% 
\begin{tabular}{ l|  l| p{7.5cm} }
\bf type & \bf subtype & \bf tags \\ \hline

\bf \em pass & cross, simple pass & accurate, not accurate, key pass, opportunity, assist, (goal)\\

\bf \em foul & & no card, yellow, red, 2nd yellow\\

\bf \em shot & & accurate, not accurate, block, opportunity, assist, (goal)\\

\bf \em duel & air duel, dribbles, tackles, ground loose ball & accurate, not accurate\\

\bf \em free kick & corner, shot, goal kick, throw in, penalty, simple kick \quad \  & accurate, not accurate, key pass, opportunity, assist, (goal)\\

\bf \em offside & & \\

\bf \em touch & acceleration, clearance, simple touch & counter attack, dangerous ball lost, missed ball, interception, opportunity, assist, (goal) \\
\hline
\end{tabular}
\caption{Event types, with their possible subtypes and tags. For further detail we remind to the Wyscout API documentation \url{https://apidocs.wyscout.com/}.}
\label{tab:events}
\end{table*}

\begin{figure*}[htb]\centering
\includegraphics[scale=0.46]{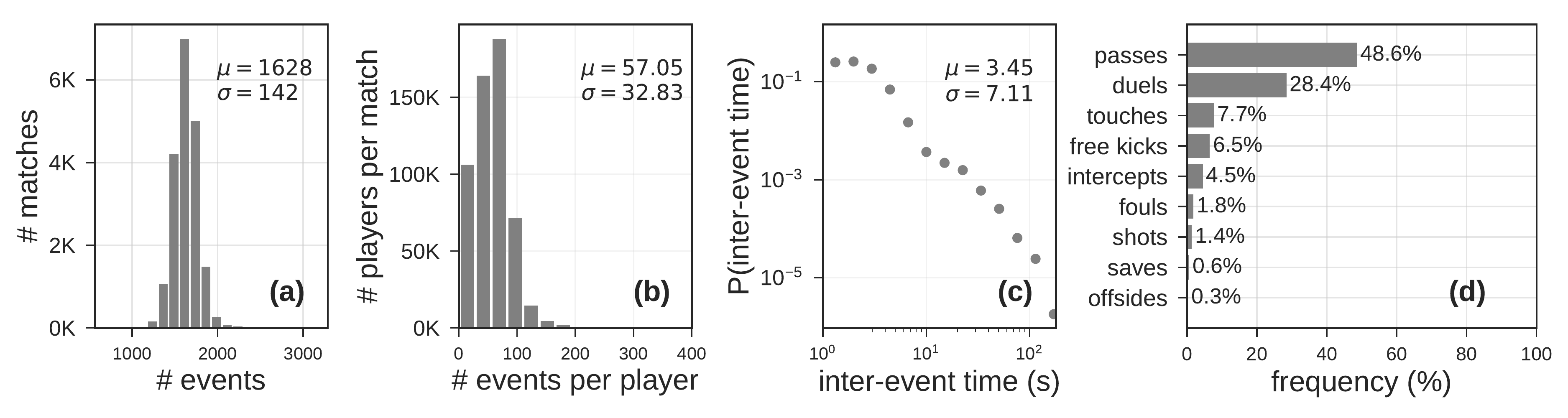}
\caption{(a) Distribution of the number of events per match ($\mu$=average, $\sigma$=st. deviation). In average, a match has 1,628 events. (b) Distribution of the number of events produced by one player per match. In average, a player generates around 57 events per match. (c) Distribution of inter-event times, defined as the time (in seconds) between two consecutive events in a match. In average, there are around three seconds between an event and the next one in a match. (d) Frequency of events per type. Passes are the most frequent event accounting for about 48\% of the events in a match.}
\label{fig:data_description}
\end{figure*}

In the Wyscout dataset a match consists of an average of about 1,600 events, and for each player there are about 57 observed events per match (Figure \ref{fig:data_description}a-b), with an average inter-time between two consecutive events of 3.45 seconds (Figure \ref{fig:data_description}c). Passes are the most frequent events, accounting for around 50\% of the total events (Figure \ref{fig:data_description}d). 
Wyscout soccer-logs adhere to a standard format for storing events collected by semi-automatic systems \cite{gudmundsson2017spatio,stein2017how,rein2016bigdata} and do not include off-ball actions. Moreover, given the existing literature on the analysis of soccer matches \cite{brooks2016developing, pappalardo2017quantifying, gyarmati2014searching, gyarmati2016analyzing, cintia2015harsh, pappalardo2017human}, we can state that the Wyscout dataset we use in our experiments is unique in the large number of events, matches and players considered, and for the length of the period of observation.

\subsection{Performance extraction}
\label{sec:performance_construction}

We compute the players' performance vectors by a two-step procedure. First, we define a feature for every possible combination of type, subtype and tag shown in Table \ref{tab:events}. For example, given the \emph{foul} type, we obtain four features: \emph{foul no card}, \emph{foul yellow}, \emph{foul red} and \emph{foul 2nd yellow}. We discard the \emph{goal} tag since we have implicitly considered the goals as the outcome of a performance during the learning phase. Nevertheless goals can be still included in the performance rating by Equation (\ref{eq:r_star}) in Section \ref{sec:rating}.
Eventually we extracted 76 features from the Wyscout soccer-logs, and normalized them in the range $[0, 1]$ in order to guarantee that all features are expressed in the same scale (see Table \ref{tab:list_features} for a list of all the features).

We tried more sophisticated features by considering the field zones where events have occurred or the fraction of the match when they have occurred, but we didn't find any significant difference w.r.t. the results presented below.

Second, we build the performance vector $\mathbf{p}_u^m$ for a player $u$ in match $m$ by counting the number of events of a given type, subtype and tag combination that player $u$ produced in $m$. For example, the number of fouls without card made by $u$ in $m$ compose the value of feature \emph{foul no card} of $u$ in $m$. 

\subsection{Role detection}
\label{sec:role_detector}
To discover roles from the Wyscout dataset we execute the role detection algorithm of Section \ref{sec:learning} by varying $k=1, \dots, 20$ and specifying $\delta_s = 0.1$, which implies that 5\% of the centers are classified as hybrids.\footnote{Experiments on the Wyscout dataset have shown that the number of hybrid centers increases linearly with $\delta_s$, from none to all centers.} We observe that $k=8$ provides the best clustering in terms of silhouette score  ($ss=0.43$) and that these results are stable across several executions of the experiment where different sets of centroids are used to initialize the $k$-means algorithm.

\begin{figure}\centering
\includegraphics[scale=0.075]{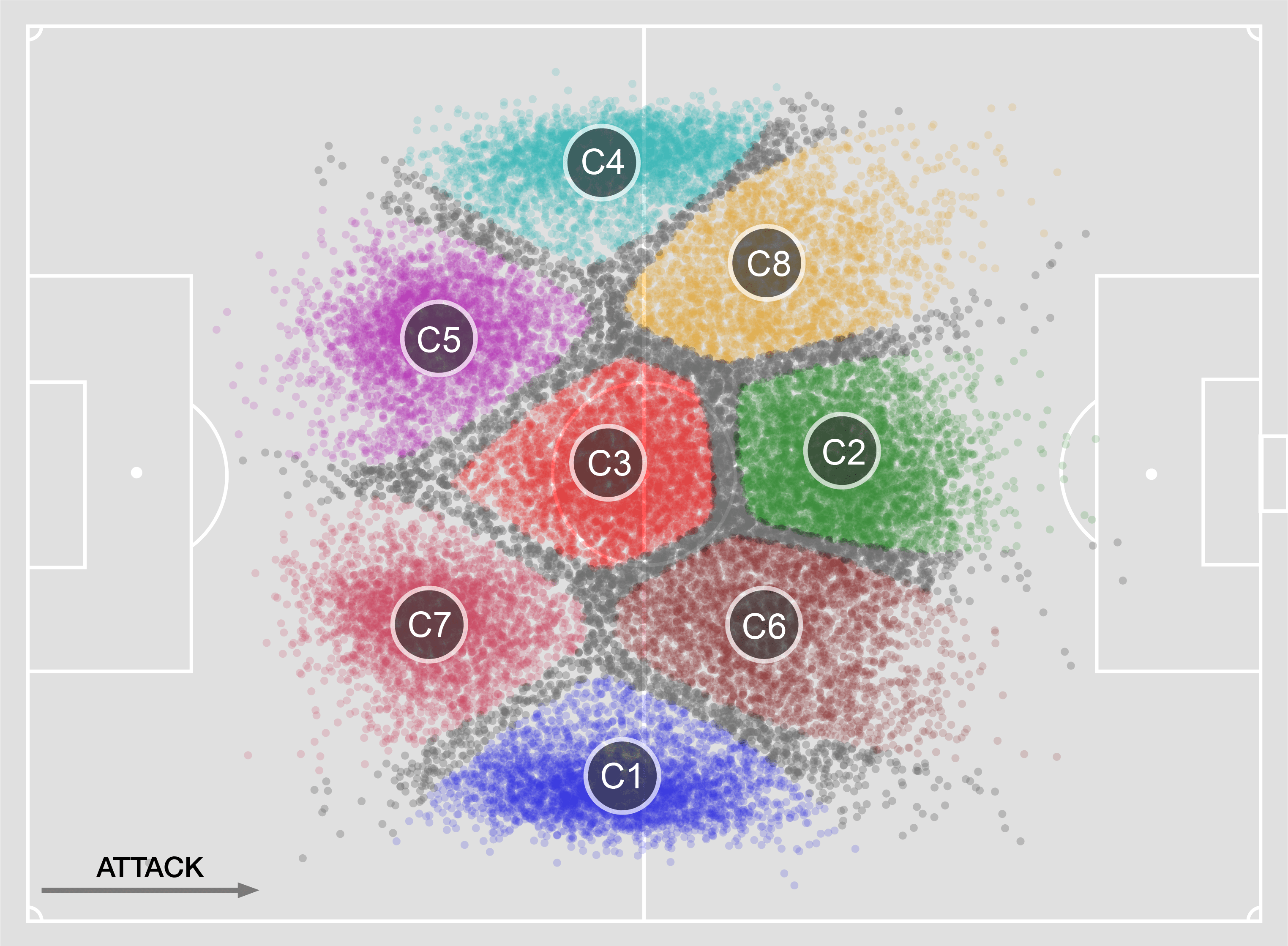}
\caption{Grouping of the centers of performance in the clusters $C_1, \dots, C_8$. Each color identifies a different cluster (role); gray points indicate hybrid centers of performance. Table \ref{tab:roles} shows an interpretation of clusters given by professional soccer scouts. Figure \ref{fig:frequency_roles} shows the frequency of a role across performances and players.}
\label{fig:silhouette_evol}
\end{figure}

Figure~\ref{fig:silhouette_evol} shows the result of the $8$-means clustering. We asked professional soccer scouts, employed by Wyscout, to provide an interpretation of the 8 clusters with terms suitable for soccer practitioners. An explanation for the clusters C1-C8, as well as a set of players typically in each role, are provided in Table \ref{tab:roles}. 

\begin{table*}[]
\def\arraystretch{1.2}
\begin{tabular}{cc|l|l}
\multicolumn{1}{l}{\textbf{cluster}} & \textbf{name}                         & \textbf{description}                                                       & \textbf{examples}      \\ \hline
C1                                   & right fielder                         & plays on the right side of the field, as a wing, back, or both & Sergi Roberto, Danilo \\
C2                                   & central forward                       & plays in the center of the field, close to the opponent's area    & Messi, Su\'arez         \\
C3                                   & central fielder                    & plays in the center of the field                                           & Kroos, Pjani\'c         \\
C4                                   & left fielder                          & plays on the left side of the field, as a wing, back, or both    & Nolito, Jordi Alba    \\
C5                                   & left central back & plays close to his own goal, preferably on the left                   & Bartra, Maguire       \\
C6                                   & right forward                         & plays on the right side of the field, close to the opponent's area      & Robben, Demb\'el\'e       \\
C7                                   & right central back                    & plays close to his own goal, preferably on the right                  & Javi Mart\'inez, Matip  \\
C8                                   & left forward                          & plays on the left side of the field, close to the opponent's area       & Neymar, Insigne \\     \hline
\end{tabular}
\caption{Interpretation of the 8 clusters detected by the role detector and examples of players assigned to each cluster. }
\label{tab:roles}
\end{table*}

It is worth to notice that, while there are 10 players in a team (excluding the goalkeeper), the clustering algorithm detected 8 roles. This means that there is at least one cluster (i.e. role) having more than one player in each team. Moreover the correspondence with classic roles is not perfect in that two players classified in two different classic roles can appear in the same cluster, and vice versa.

\begin{figure}
    \centering
    \includegraphics[scale=0.55]{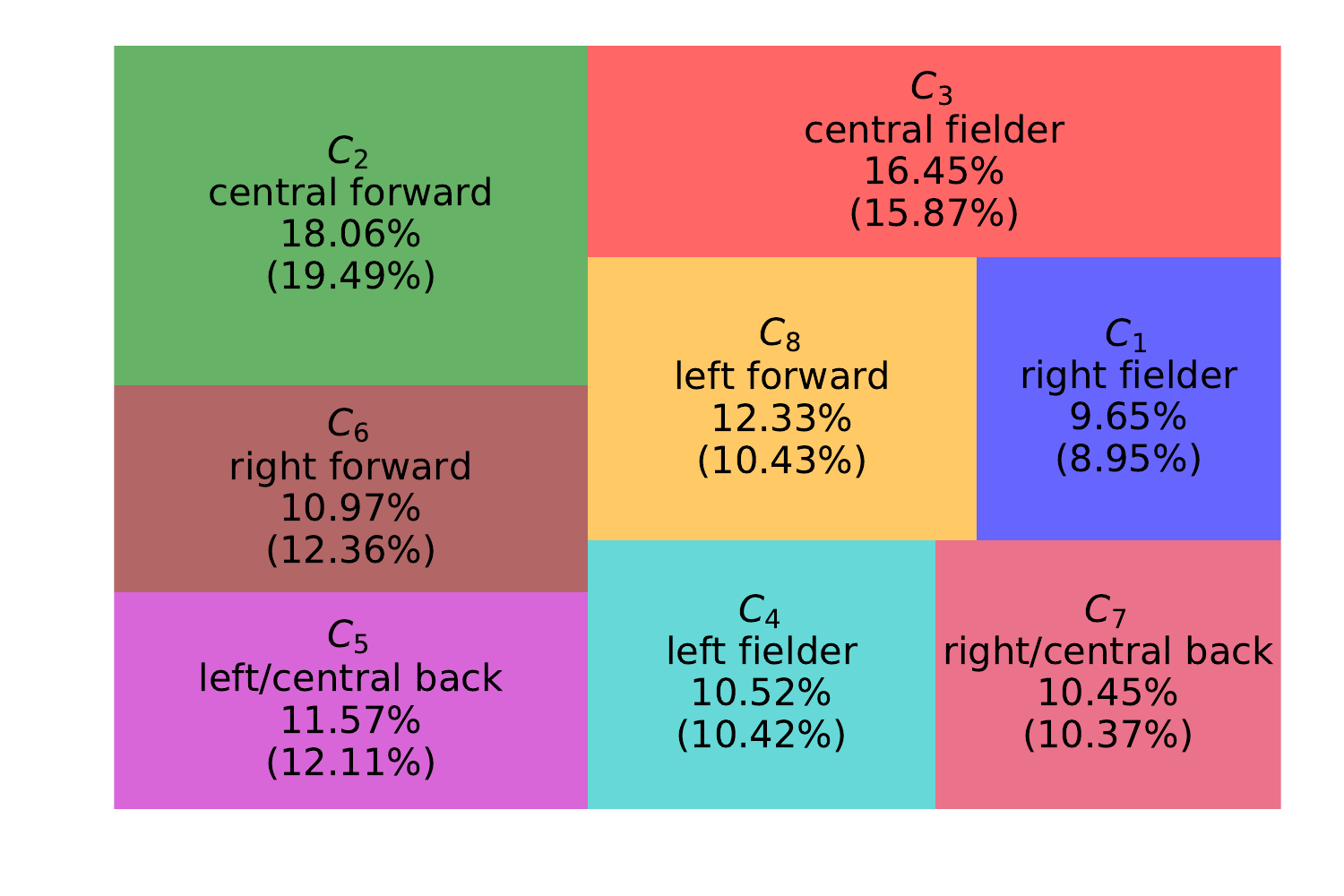}
    \caption{Distribution of the 8 roles discovered by the role detector across performances and players (in parenthesis) within our dataset. Each player is assigned to the role he covers most frequently during the available seasons.}
    \label{fig:frequency_roles}
\end{figure}

Figure \ref{fig:frequency_roles} shows how the performances and the players are distributed among the detected roles, where each player is assigned to the role he covers most frequently during the matches of the available seasons. We find that role $C_2$ (central forward) is the most common role covering 18\% of performances and 19\% of players, followed by role $C_3$ (central fielder) covering 16\% of performances and 15\% of players. All other roles are almost equally populated.

\subsection{Feature weighting}
\label{sec:feature_weighting}
As discussed in Section~\ref{sec:learning}, {\sf PlayeRank} turns the problem of estimating the $76$ feature weights into a classification problem between a team performance vector and a match outcome. We instantiate this problem by creating, for each match $m$, two examples $\mathbf{p}_{T_1}^m$ and $\mathbf{p}_{T_2}^m$, that correspond to the performance vectors of two playing teams $T_1$ and $T_2$, and the match outcome label $o_T^m$ that is 1 if a team wins and 0 otherwise. The resulting dataset consists of $19,619$ examples, 80\% of which are used to train a Linear Support Vector Machine (SVM). We have selected the cost parameters that had the maximum average Area Under the Receiver Operating Characteristic Curve (AUC) on a 5-fold cross validation.
We validate SVM on the remaining 20\% of the examples, finding an $AUC {=} 0.89$ ($F1{=}0.81$, accuracy${=}0.82$), significantly better than a classifier which always predicts the most frequent outcome (i.e., non-victory, $AUC{=}0.50$, $F1{=}0.48$, accuracy${=}0.62$) and a classifier which chooses the label at random based on the distribution of victories and non-victories ($AUC{=}0.50$, $F1{=}0.53$, accuracy${=}0.53$). 
We also experimented with different labelling of $o_T^m$ by defining either $o_T^m {=} 0$ in the case of defeat and $o_T^m {=} 1$ otherwise, or by defining a ternary classification problem where $o_T^m {=} 1$ indicates a victory, $o_T^m {=} 0$ a defeat and $o_T^m {=} 2$ a draw. In all these cases we did not find any significant difference in the feature weights described below, so that we chose to deploy the binary classification problem above.

\begin{figure}[htb]\centering
\includegraphics[scale=0.5]{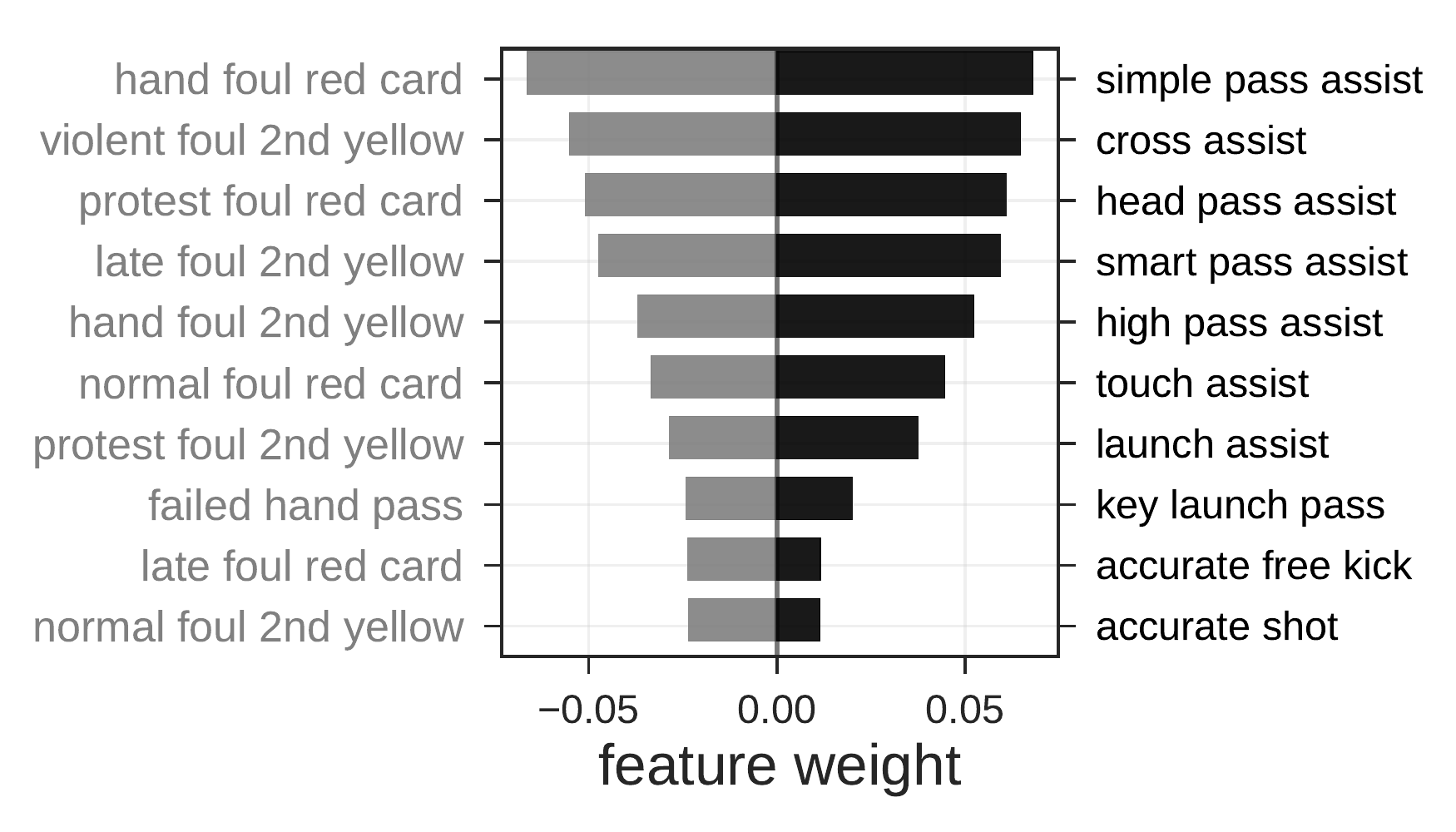}
\caption{Top-10 (black bars, on the right) and bottom-10 (gray bars, on the left) features according to the value of the weights extracted from the SVM model trained on all competitions together.}
\label{fig:importances}
\end{figure}

Figure \ref{fig:importances} shows the top-10 (black bars) and the bottom-10 (grey bars) feature weights $\mathbf{w} = [w_1, \dots, w_n]$ resulting from SVM. We find that assist-based features are the most important ones, followed by the number of key passes and the accuracy of shots. In contrast, getting a red/yellow card gets a strong negative weight, especially for hand and violent fouls. It is interesting to notice that, though these choices are pretty natural for who is skilled in soccer-player evaluations, {\sf PlayeRank} derived them automatically by just looking at the massive soccer-logs provided by Wyscout.

\begin{figure*}[htb]\centering
\includegraphics[scale=0.55]{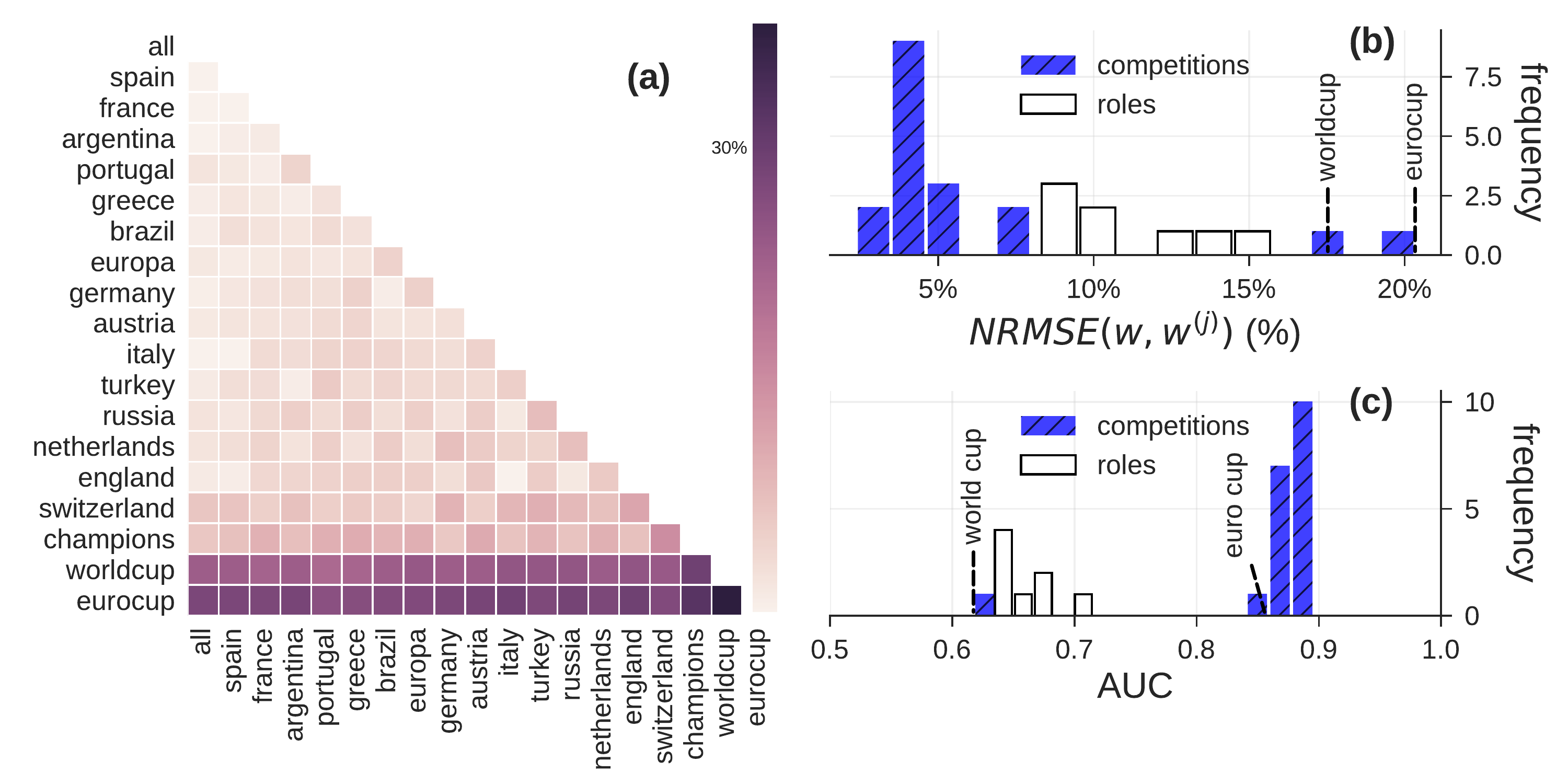}
\caption{(a) Heatmap indicating the Normalized Root Mean Squared Error (NRMSE) between the set of feature weights $\mathbf{w}^{(j)}$ of each competition $j$ and the overall set of feature weights $\mathbf{w}$. The error is higher for UEFA Euro Cup 2016 and the FIFA World Cup 2018. (b) Distribution of $NRMSE(\mathbf{w}, \mathbf{w}^{(j)})$, expressed in percentage, indicating the normalized error between $\mathbf{w}$ and a competition's set of weights (blue bars) and a role's set of weights (white bars). The error is for 16 out of 18 competitions below 7\%, it is around 17\% and 20\% for EUFA Euro Cup 2016 and FIFA World Cup 2018, respectively. For role-based weights, the error is between 8\% and 15\%, higher than the majority of the competitions' weights but lower than the weights of UEFA Euro Cup 2016 and FIFA World Cup 2018. (c) Distribution of AUC of the SVM models trained on the 18 competitions separately (blue bars) and the 8 roles separately (white bars). The average AUC of competition-based models is 0.86 and the AUC of the SVM model trained on the FIFA World Cup 2018 is lower than all other models. For role-based models, the AUC is closer to the FIFA World Cup 2018's AUC than to the AUC of the other competitions' models.}
\label{fig:nrmse}
\end{figure*}

For the sake of completeness of our experimental results, we also repeated the classification task separately: (i) competition by competition, i.e., we created 18 SVMs each one trained on the matches of one competition only; (ii) role by role, i.e., we created 8 SVMs each one trained on the examples created from players tagged with one role only (Section \ref{sec:role_detector}).
\paragraph{Competition-based weights.} We extracted for each of the 18 competitions the corresponding set of weights $\mathbf{w}^{(j)} = [w_1^{(j)}, \dots, w_n^{(j)}]$ ($j=1, \dots, 18$) and quantified the difference between the weights $\mathbf{w}$ extracted from all competitions and the $\mathbf{w}^{(j)}$s via the Normalized Root-Mean-Square Error:
\begin{equation}
NRMSE(\mathbf{w}, \mathbf{w}^{(j)}) = \frac{\sqrt[]{\frac{1}{n} \sum_{i=1}^n (w_i - w_i^{(j)})^2 }}{\max{\mathbf{w}}- \min{\mathbf{w}}}
\end{equation}
where $\max{\mathbf{w}}$ and $\min{\mathbf{w}}$ are the maximum and the minimum weights in $\mathbf{w}$, respectively. We found that the average $NRMSE$ is around 6\% and that 16 out of 18 competitions have $NRMSE < 7\%$ (Figure \ref{fig:nrmse}b), indicating that the difference between any $\mathbf{w}^{(j)}$ and $\mathbf{w}$ is small and hence the relation between team performance and match outcome is in most of the cases independent of the specific competition for clubs considered. Only for competitions involving national teams, such as UEFA Euro Cup 2016 and FIFA World Cup 2018, the $NRMSE$ is higher, 17\% and 20\% respectively (Figure \ref{fig:nrmse}b). This can be due either to the fact that: (i) these two competitions have a few matches (51 and 64 respectively, see Table \ref{tab:competitions}) or that (ii) while all the other competitions refer to soccer clubs, UEFA Euro Cup 2016 and FIFA World Cup 2018 are competitions for national teams, which are generally more unpredictable \cite{cintia2015harsh, cintia2015network}. Figure \ref{fig:nrmse}c indicates that the accuracy of the SVM model trained on the 64 matches of the FIFA World Cup 2018 is lower than the accuracy of the other models, suggesting that the number of matches in a competition influences the accuracy of the model. However, the accuracy of the SVM model trained on the UEFA Euro Cup 2016 is close to the accuracy of all other models, suggesting that the difference in the weights can be also due to the specific nature of the competition.  

\paragraph{Role-based weights.}
We repeated the classification task separately role by role by aggregating the players' feature role by role.
We found that: (i) the accuracy of the SVM models trained on the roles separately are lower than the accuracy of the models trained on the competitions, though the role-based model's accuracy is still higher than the model trained on the FIFA World Cup 2018  (Figure \ref{fig:nrmse}c); (ii) the $NRMSE$ between each role's set of weights and the set of weights trained on all competitions together is lower than 15\% (Figure \ref{fig:nrmse}b). This indicates that there is a small variation between the competition-based and the role-based sets of weights. For this reason, we will just use $\mathbf{w}$ (i.e., the set of weights computed at match level including all competitions) in the computation of the ratings in the following sections.

\subsection{Player ratings and rankings}
\label{sec:ratings}

\begin{figure}[htb]
    \centering
    \includegraphics[scale=0.5]{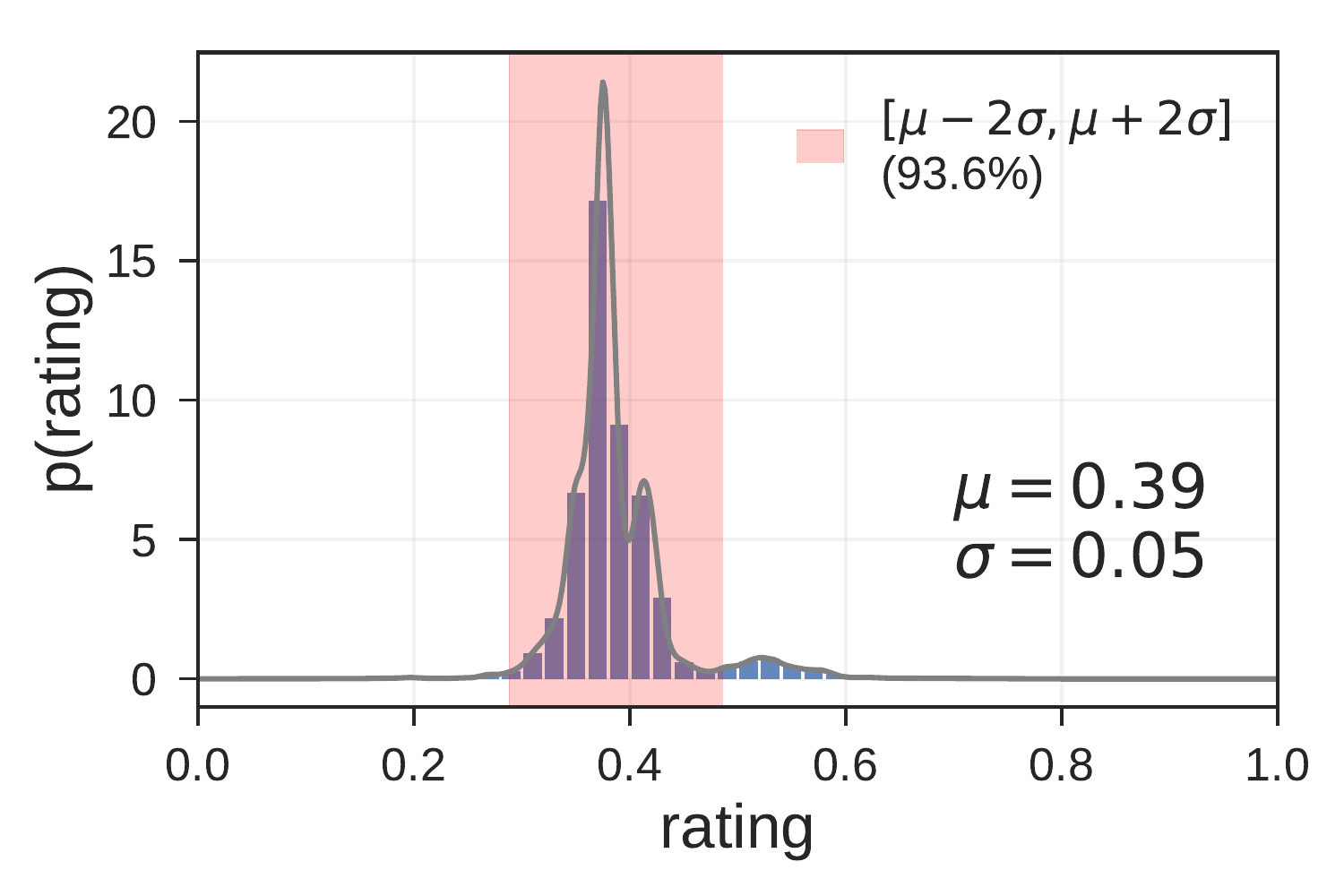}
    \caption{Distribution of performance ratings. It is strongly peaked around the average $\mu{=}0.39$, while outliers are rare. Most of the ratings ($\approx$ 94\%) are within the range $[\mu - 2\sigma, \mu + 2\sigma]$ ($\sigma$ is the standard deviation).}
    \label{fig:distribution_ratings}
\end{figure}

\begin{figure*}
    \centering
    \includegraphics[scale=0.39]{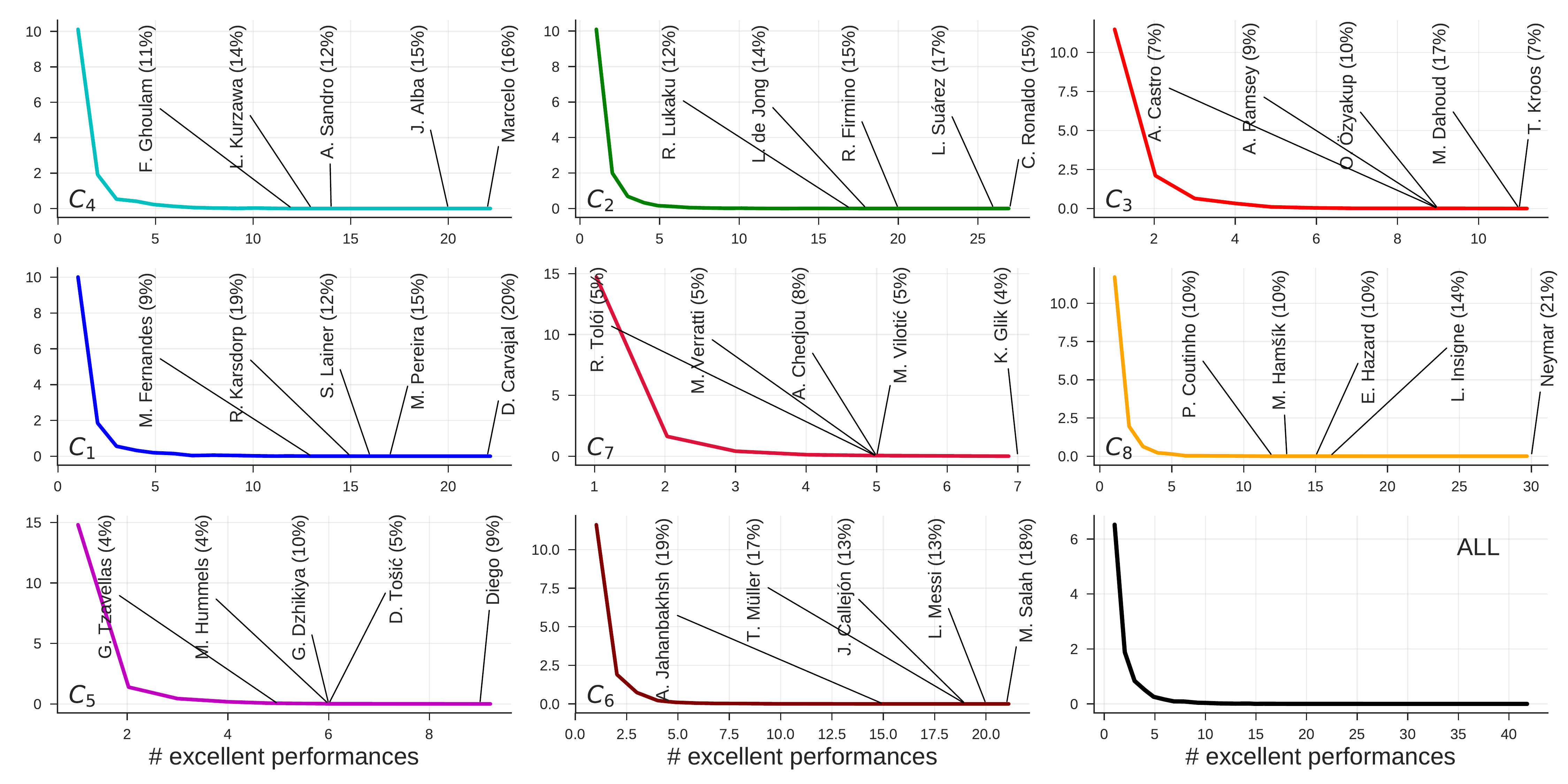}
    \caption{Distribution of the number of excellent performances (i.e., $r(u, m) > \mu + 2\sigma$) per player, for each role ($C_1$, \dots, $C_8$) and for all roles together ($ALL$). The name of the players who achieve the top 5 performances in each role is showed. The y axis indicates the probability density function of the number of excellent performances.}
    \label{fig:power_laws}
\end{figure*}

Given $\mathbf{w}$, we compute the performance rating $r(u, m)$ for each player $u$ in each match $m$ and then explore their distribution. 
As Figure \ref{fig:distribution_ratings} shows, the distribution is strongly peaked around its average ($\mu=0.39$), indicating that ``outlier'' performances (i.e.,  $r(u, m) \not\in [\mu - 2\sigma, \mu + 2\sigma]$, $\sigma$ is the standard deviation) are rare. 
In particular,  excellent performances (i.e., $r(u, m) > \mu + 2\sigma$), accounting for just 5\% of the total, are unevenly distributed across the players. Indeed, the distribution of the number of excellent performances per player is long-tailed (Figure \ref{fig:power_laws}, ALL): while the majority of players achieve a few excellent performances, a tiny fraction of players achieve up to 40 excellent performances during the five years. This trend is observed also when we split performances by the player's role, highlighting the presence of a general pattern (Figure \ref{fig:power_laws}, $C_1$-$C_8$).

\begin{figure*}
    \centering
    \includegraphics[scale=0.38]{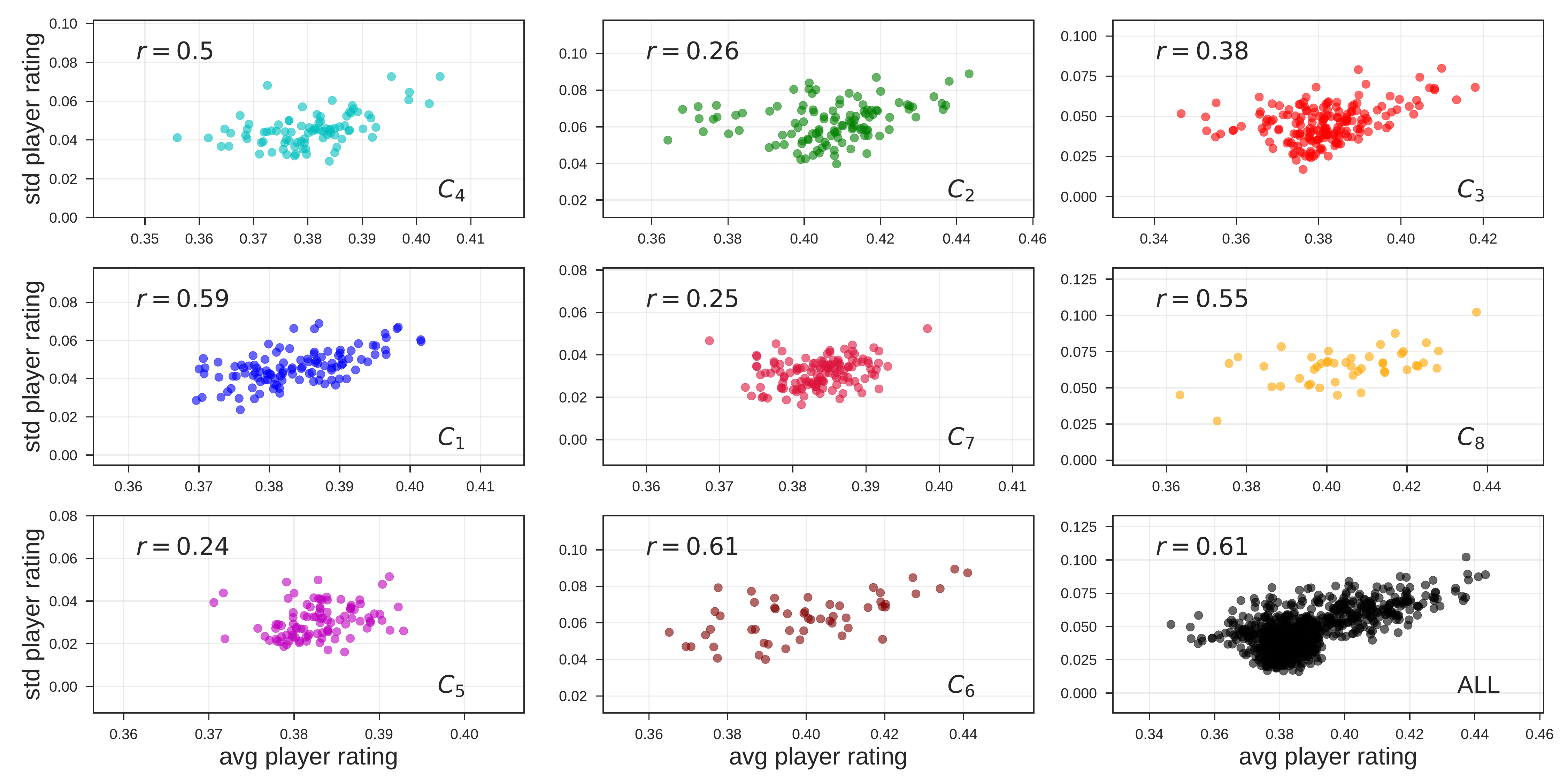}
    \caption{Correlation between a player's average performance rating and his standard deviation, for each role ($C_1$, \dots, $C_8$) and all roles together ($ALL$). Here $r$ indicates the Pearson correlation coefficient.}
    \label{fig:mean_vs_std}
\end{figure*}

As an example, let us consider all performances of role $C_6$ (left forward): most of the players achieve excellence just once, while a few players achieve as many as 30 (Neymar, 21\% of his performances), 16 (L. Insigne, 14\%) and 15 (E. Hazard, 10\%) excellent performances. 
Moreover, we find that a correlation exists between a player's average performance rating and the variability of his ratings (Figure \ref{fig:mean_vs_std}): the stronger a player is (i.e., the higher his average performance rating), the more variable his performance ratings are (i.e., the higher is the standard deviation of his ratings). 
In other words, the best players do not play excellence in every match, they just achieve excellence more frequently than the other players. 
Taken together, Figures \ref{fig:distribution_ratings}, \ref{fig:power_laws} and \ref{fig:mean_vs_std} indicate that: (i) excellent performances are rare ($\approx$5\% of the total); (ii) just 11\% of the players achieve excellent performances at least once; (iii) while a small set of players repeatedly achieve excellence, all other players do a few times, suggesting that the best players do not always play excellence but they just achieve it more frequently; (iv) excellent performances are at most 21\% (Neymar) and on average 9\% of all the performance of players who reach excellence at least once.

\begin{figure*}[h]
    \centering
    \includegraphics[scale=0.40]{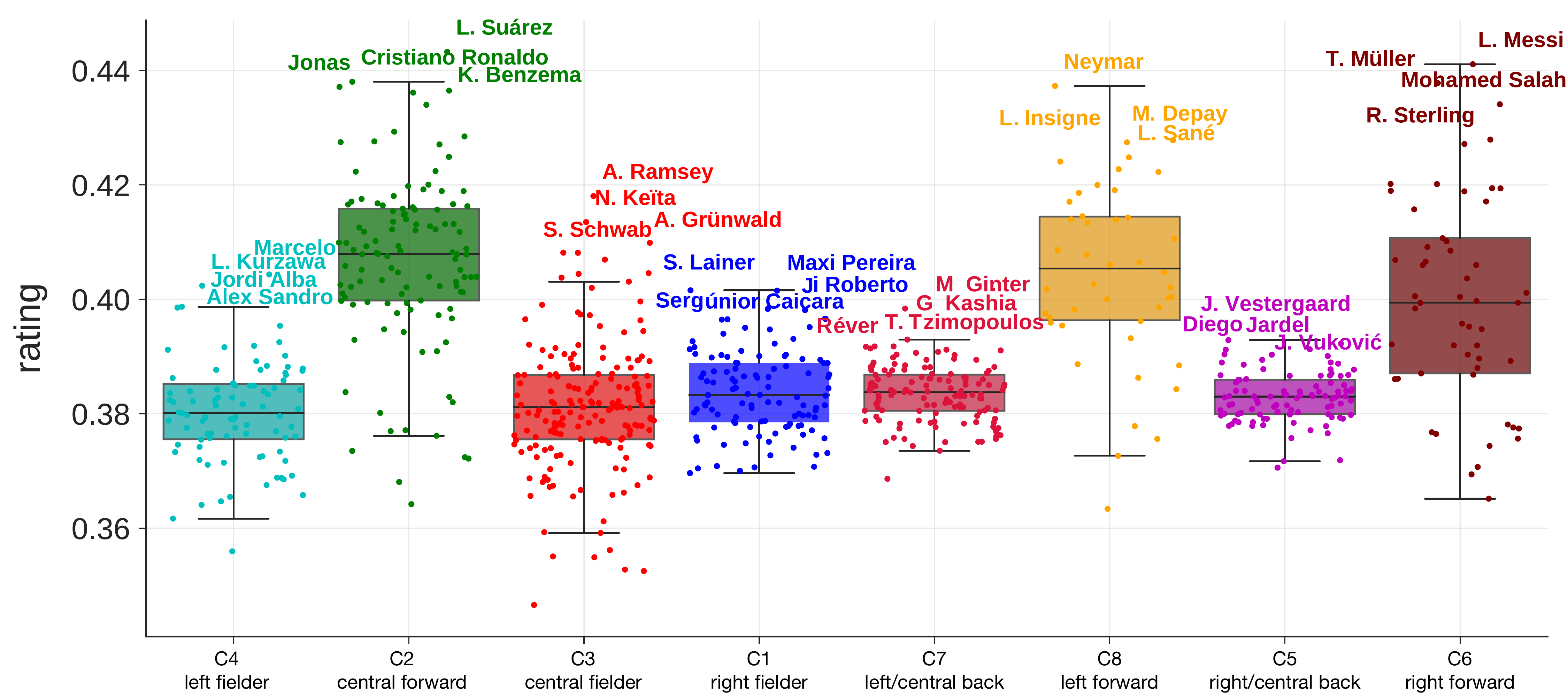}
    \caption{Distribution of player ratings per role. Each boxplot represents a cluster (role) and each point (circle) indicates a player's rating, computed across all the performances in the last four seasons of the 18 competitions. The points are jittered by adding random noise to the x-axis value to help visualization. For each cluster, the players' name at the top of the corresponding role-based rankings are shown.}
    \label{fig:ranking}
\end{figure*}

By aggregating the performance rating of each player $u$ over the whole series $M$ of matches, we compute the player rating $\overline{r}(u, M)$. Figure~\ref{fig:ranking} visualizes the distribution of these player ratings by grouping the players on the x-axis according to their roles. We recall that we are assigning a player to a role if he plays at least 40\% of the matches in that role, meaning that a player may be assigned to at most two roles among the 8 roles detected. We observe a different distribution of ratings according to the players' roles, both in terms of range of values and their concentration. This fully justifies the design of the role detection module in the {\sf PlayeRank} framework. In fact we notice that the top-ranked player of cluster $C_4$, Marcelo, gets a player rating which is below the average of the ratings of clusters $C_6$ or $C_8$ (Figure \ref{fig:ranking}).

Table \ref{tab:cluster_ranking} reports the top-10 players grouped by the 8 roles. Although {\sf PlayeRank} is fully data-driven, it is able to place the most popular players at the top of some ranking. For example, Lionel Messi (Barcelona) is the best player in cluster $C_6$ (see Figure~\ref{fig:silhouette_evol}), followed by other renowned players such as Thomas M\"{u}ller (Bayern Munich) and Mohamed Salah (Liverpool). Instead, the best player in cluster $C_2$ (central forward) is Lu\'{i}s Su\'{a}rez (Barcelona), preceding Cristiano Ronaldo (Juventus), Jonas (Benfica) and Benzema (Real Madrid). Other renowned players are at the top of their role's ranking, such as Neymar (PSG, cluster $C_8$, left forward) and Marcelo (Real Madrid, cluster $C_4$, left fielder). 
%Table \ref{tab:cluster_ranking} also shows the utility of the soft-clustering procedure: some players are assigned to multiple rankings, indicating that they can play in different roles, highlighting their versatility. For example, this is the case of Cristiano Ronaldo, which appears both in clusters $C_2$ and $C_6$. On the other hand, our classification in 8 roles allows us to distinguish players that were classified together by the classic three-way classification in defenders-midfielder-forwards. This is the case of Messi and De Bruyne which are in different roles according to the traditional classification (because the former is forward, the latter is midfielder), whereas they appear in the same cluster according to {\sf PlayeRank} given how they play on the field.

\begin{figure}[H]
    \centering
    \includegraphics[scale=0.5]{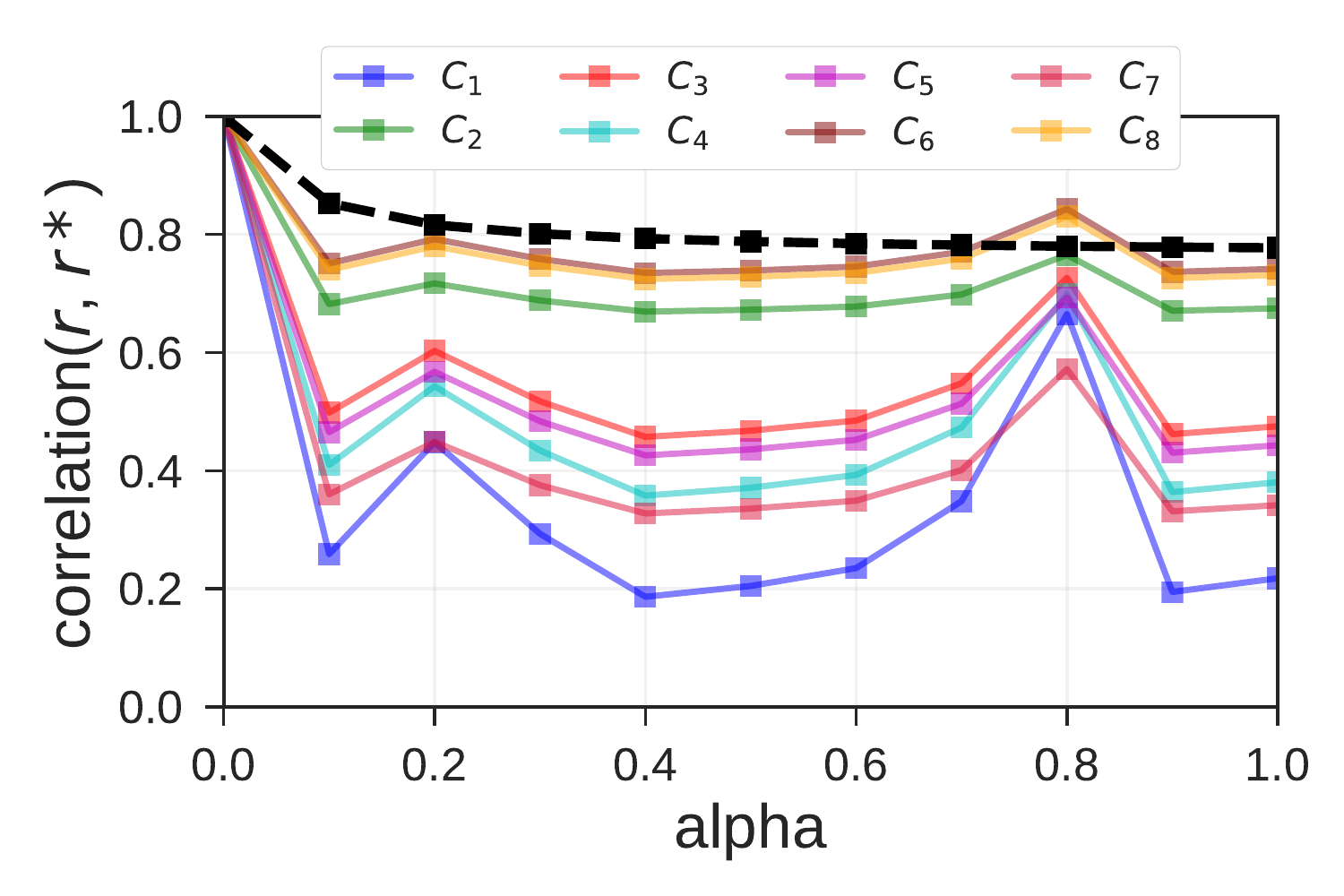}
    \caption{Correlation between player ratings and adjusted-player ratings as  $\alpha$ varies in the range $[0, 1]$. The dashed curve refers to all players together, the solid to the 8 roles.}
    \label{fig:ratingvsadjusted}
\end{figure}

What it is surprising in these role-based rankings is that they have been derived by {\sf PlayeRank} without considering the number of goals scored by players when building the performance vector.
Actually, we observe that in general the goal-adjusted ranking $\overline{r}^*(u, M)$ is consistent with $\overline{r}(u, M)$ for all values of $\alpha$ (Eq. \ref{eq:r_star}): as the black dashed curve in Figure \ref{fig:ratingvsadjusted} shows, the correlation between the player rating and the adjusted-player rating slightly decreases with $\alpha$, with values that are in general $\geq 0.8$.  
However, when investigating how the correlation changes with $\alpha$ role by role, we find that while offensive-oriented roles like $C_2$ (central forward), $C_6$ (right forward) and $C_8$ (left forward) show in general high correlations between those ratings ($\in [0.65, 0.85]$), roles $C_3$ (central fielder), $C_5$ (left central back), $C_4$ (left fielder) shows moderate correlations ($\in [0.4, 0.75]$), while role $C_1$ (right fielder) shows low correlation ($\in [0.2, 0.65]$). This result suggests that the player rating of offensive players is not much influenced by the number of goals scored, presumably because they are already associated with events related to scoring. 

\begin{table*}[]
    \centering
    \begin{tabular}{r | r | l | l || r | r | l | l || r | r | l | l |}
\hline
  &  \bf \em r &            \bf player &                \bf club &   &  \bf \em r &              \bf player &                 \bf club &   &  \bf \em r &            \bf player &                \bf club \\
\hline
   \multirow{10}{*}{\rotatebox[origin=c]{90}{\textbf{ cluster 4} - left fielder}}       &    .404 &          Marcelo &          R. Madrid &    \multirow{10}{*}{\rotatebox[origin=c]{90}{\textbf{cluster 2} - central foward}}       &    .404 &          L. Su\'arez &             Barcelona &      \multirow{10}{*}{\rotatebox[origin=c]{90}{\textbf{cluster 3} - central fielder }}     &    .404 &          A. Ramsey &             Arsenal \\
          &    .402 &       L. Kurzawa &                  PSG &           &    .402 &  C. Ronaldo &              Juventus &           &    .402 &           N. Ke\"ita &          RB Leipzig \\
          &    .399 &      Alex Sandro &             Juventus &           &    .399 &              Jonas &               Benfica &           &    .399 &        A. Gr\"unwald &        A. Wien \\
          &    .399 &       Jordi Alba &            Barcelona &           &    .399 &         K. Benzema &           R. Madrid &           &    .399 &          S. Schwab &          R. Wien \\
          &    .395 &       J. Willems &  Eintracht F. &           &    .395 &         D. Mertens &                Napoli &           &    .395 &     J. van Overeem &                  AZ \\
          &    .393 &         D. Alaba &       Bayern M. &           &    .393 &         L. de Jong &                   PSV &           &    .393 &         O. \"Ozyakup &            Be\c{s}ikta\c{s} \\
          &    .392 &    Marcos Alonso &              Chelsea &           &    .392 &          S. Ag\"uero &       Man City &           &    .392 &         A. Dzagoev &         CSKA M. \\
          &    .392 &        B. Davies &    Tottenham &           &    .392 &      Son Heung-Min &     Tottenham &           &    .392 &         C. Tolisso &      Bayern M. \\
          &    .391 &      D. Kombarov &       Spartak M. &           &    .391 &            D. Alli &     Tottenham  &           &    .391 &      R. Nainggolan &                Roma \\
          &    .390 &        J. Brenet &                  PSV &           &    .390 &          A. Dzyuba &                 Zenit &           &    .390 &         B. N'Diaye &          Stoke City \\
          \hline \hline
          
          \multirow{10}{*}{\rotatebox[origin=c]{90}{\textbf {cluster 1} - right fielder}} &    .402 &        S. Lainer &             Salzburg &           \multirow{10}{*}{\rotatebox[origin=c]{90}{\textbf {cluster 7} - central back}} &    .402 &          M. Ginter &   Borussia M. &           \multirow{10}{*}{\rotatebox[origin=c]{90}{\textbf{ cluster 8} - left forward}} &    .402 &             Neymar &                 PSG \\
          &    .402 &     Maxi Pereira &                Porto &           &    .402 &          G. Kashia &               Vitesse &           &    .402 &           M. Depay &  Olympique L. \\
          &    .398 &    Sergi Roberto &            Barcelona &           &    .398 &              R\'ever &              Flamengo &           &    .398 &         L. Insigne &              Napoli \\
          &    .398 &   J\'unior Cai\c{c}ara &  Istanbul B. &           &    .398 &     T. Tzimopoulos &          PAS &           &    .398 &            L. San\'e &     Man City \\
          &    .397 &  Daniel Carvajal &          R. Madrid &           &    .397 &           M. Yumlu &  Akhisar &           &    .397 &          M. Ham\v{s}\'ik &              Napoli \\
          &    .397 &  L. De Silvestri &               Torino &           &    .397 &             Hilton &           Montpellier &           &    .397 &          M. Dabbur &            Salzburg \\
          &    .397 &  Ricardo Pereira &       Leicester &           &    .397 &    T. Alderweireld &     Tottenham  &           &    .397 &          E. Hazard &             Chelsea \\
          &    .396 &     D. Caligiuri &           Schalke &           &    .396 &        Bruno Silva &              Cruzeiro &           &    .396 &  P. Coutinho &           Barcelona \\
          &    .395 &        N. Skubic &            Konyaspor &           &    .395 &           Y. Ayhan &        Osmanlıspor &           &    .395 &         I. Peri\v{s}i\'c &      Inter \\
          &    .395 &        S. Widmer &              Udinese &           &    .395 &         J. Schunke &           Estudiantes &           &    .395 &               Isco &         R. Madrid \\
          \hline \hline
          
          \multirow{10}{*}{\rotatebox[origin=c]{90}{\textbf{ cluster 5} - central back}} &    .393 &   J. Vestergaard &          Southampton &           \multirow{10}{*}{\rotatebox[origin=c]{90}{\textbf{ cluster 6} - right forward}} &    .393 &           L. Messi &             Barcelona            \\
          &    .392 &           Jardel &              Benfica &           &    .392 &          T. M\"uller &        Bayern M.          \\
          &    .391 &       J. Vukovi\'c &        Hellas Verona &           &    .391 &      M. Salah &             Liverpool                   \\
          &    .391 &            Diego &          Antalyaspor &           &    .391 &        R. Sterling &       Man City            \\
          &    .390 &       Ra\'ul Silva &       Sporting Braga &           &    .390 &            G. Bale &           R. Madrid            \\
          &    .390 &        D. Siovas &              Legan\'es &           &    .390 &            S. Man\'e  &            Liverpool           \\
          &    .390 &       M. Hummels &       Bayern M. &           &    .390 &       K. Bellarabi &      B. Leverkusen            \\
          &    .389 &          C. Lema &             Belgrano &           &    .389 &          B. Traor\'e &    Olympique L.            \\
          &    .389 &        L. Perrin &        Saint-\'Etienne &           &    .389 &     Gelson Martins &       A. Madrid           \\
          &    .389 &   S. Ignashevich & CSKA M.                     &           &    .389 &        A. Candreva &        Inter                                 \\
\cline{1-8}
\end{tabular}
\caption{Top-10 players in each role-based ranking, with the corresponding player rating ($r$) computed across the last four seasons of the 18 competitions. The club indicated in the table is the one the player played with at the end of 2018.}
\label{tab:cluster_ranking}
\end{table*}

\section{Validation of PlayeRank}
\label{sec:validation}

Existing player ranking approaches report judgments that consist mainly of informal interpretations based on some simplistic metrics (e.g., market value or goals scored \cite{torgler2007shapes,stanojevic2016towards,brooks2016developing}). It is important instead to evaluate the goodness of ranking and performance evaluation algorithms in a quantitative manner, through the help of human experts as done for example for the evaluation of recommender systems in information retrieval. 

We validated {\sf PlayeRank}  by creating and submitting a survey to three professional soccer talent scouts, employed by Wyscout, hence particularly skilled at evaluating and comparing soccer players. Our survey consisted of a set of pairs of players randomly generated by a two-step procedure, defined as follows. First, we randomly selected 35\% of the players in the dataset. Second, for each selected player $u$ we cyclically iterated over the ranges $[1,10]$, $[11,20]$ and $[21, \infty]$ and selected one value, say $x$, for each of these ranges, and then picked the player being $x$ positions above $u$ and the one being $x$ positions below $u$ in the role-based ranking (if they exist). This generated a set $P$ of 211 pairs involving 202 distinct players. 

For each pair $(u_1, u_2) \in P$, each soccer scout was asked to select the best player between $u_1$ and $u_2$, or to specify that the two players were equally valuable. For each such pair, we also computed the best player according to {\sf PlayeRank} by declaring $u_1$ stronger than $u_2$ if $u_1$ precedes $u_2$ in the ranking. We then discarded from $P$ all pairs for which  there is not a majority among the evaluations of the soccer experts: namely, either all experts expressed equality or two experts disagreed in judging the best player and the third one expressed equality. As a result of this process, we discarded 8\% of $P$'s pairs.

\begin{figure}[htb]\centering
\includegraphics[scale=0.455]{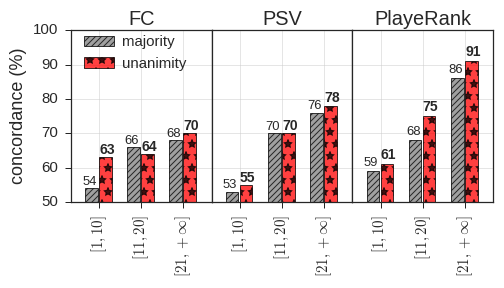}
   \caption{Majority (gray bars) and unanimity (red) concordance between Flow Centrality and the scouts (left), the PSV and the scouts (center), PlayeRank and the scouts (right).}
   \label{fig:evaluation}
\end{figure}

Over the remaining $P$'s pairs, we investigated two types of concordance among the scouts' evaluations: (i) the \emph{majority concordance} $c_{\mbox{\footnotesize maj}}$ defined as the fraction of the pairs for which {\sf PlayeRank} agrees with at least two scouts; (ii) the \emph{unanimous concordance} $c_{\mbox{\footnotesize una}}$ defined as the fraction of pairs for which the scouts' choices are unanimous and {\sf PlayeRank} agrees with them. We found that  $c_{\mbox{\footnotesize maj}} = 68\%$ and $c_{\mbox{\footnotesize una}} = 74\%$, indicating that {\sf PlayeRank} has in general a good agreement with the soccer scouts, compared to the random choice (for which $c_{\mbox{\footnotesize maj}} {=} c_{\mbox{\footnotesize una}} {=} 50\%$). Figure \ref{fig:evaluation} offers a more detailed view on the results of the survey by specializing $c_{\mbox{\footnotesize maj}}$ and $c_{\mbox{\footnotesize una}}$ on the three ranges of ranking differences: $[1,10], [11,20], [21, \infty]$. The bars show a clear and strong correlation between the concordance among scouts' evaluations (per majority or unanimity) and the difference between the positions in the ranking of the checked pairs of players: when the ranking difference is $\leq 10$ it is $c_{\mbox{\footnotesize maj}} = 59\%$ and $c_{\mbox{\footnotesize abs}} = 61\%$; for larger and larger ranking differences, {\sf PlayeRank} achieves a much higher concordance with experts which is up to $c_{\mbox{\footnotesize maj}} = 86\%$ and $c_{\mbox{\footnotesize abs}} = 91\%$ when the ranking difference is $\geq 20$. Clearly, the disagreement between {\sf PlayeRank} with the soccer scouts is less significant when the players are close in the ranking (i.e., their distance < 10). Indeed, the comparison between soccer players is a well-known difficult problem as witnessed by the significant increase in the fraction of unanimous answers by the scouts, which goes from a low 58\% in the range $[1,10]$ to a reasonable 71\% in the range $[21,+\infty]$. This {\em a fortiori} highlights the robustness of {\sf PlayeRank}: the scouts disagreement decreases as pairs of players are farther and farther in the ranking provided by {\sf PlayeRank}.

As a final investigation, we compared {\sf PlayeRank} with the Flow Centrality (FC) \cite{duch2010quantifying} and the PSV \cite{brooks2016developing} metrics, which constitute the current state-of-the-art in soccer-players ranking (see Section \ref{sec:related}). These metrics are somewhat {\em mono-dimensional} because they exploit just passes or shots to derive the final ranking. Figure \ref{fig:evaluation} (right) shows the results obtained by FC and PSV over our set of players' pairs evaluated by the three Wyscout experts. It is evident that FC and PSV achieve significantly lower concordance than {\sf PlayeRank} with the experts: for PSV, the majority concordance ranges from 53\% to 76\%, while the unanimity concordance ranges from 55\% to 78\%; for FC the majority concordance ranges from 54\% to 68\%, while the unanimity concordance ranges from 63\% to 70\%. So that {\sf PlayeRank} introduces an improvement which is up to 16\% (relative) and 13\% (absolute) with respect to PSV, and and an improvement of 30\% (relative) and 21\% (absolute) with respect to FC. 

\section{Applications}
To demonstrate its usefulness, in this section we show two examples of analytical services that can be designed using {\sf PlayeRank}: the retrieval of players in a database of soccer-logs and the computation of the players' versatility.

\subsection{Retrieval of players}
\label{sec:retrieving_results}

One of the most useful applications of {\sf PlayeRank} is searching players in a soccer-logs database. The search is driven by a query formulated in terms of a suitable {\em query language} that considers the events occurring during a match and their position on the field. Since we do not want to enter in the formal definition of the {\em full} query language, which is beyond the scope of this paper, we concentrate here only on its specialties that are the most interesting {\em algorithmically} for the issues we have discussed in this paper. 

We propose the efficient solution of a {\em spatial query} over the soccer-field zones which possibly span more roles and have geometric forms that differ from the ones identified by the role detector. 
We assume a tessellation of the soccer field into $h$ zones of equal size $z_1, \dots, z_h$. The query is modeled as a vector $Q = [q_1, \dots, q_h]$ in which $q_i$ expresses {\em how much relevant} is the presence of the searched player in zone $z_i$. Similarly, player $u$ is modeled as a vector $V_u = [u_1, \dots, u_h]$ in which $u_i$ expresses {\em how much inclined} is player $u$ to play in zone $z_i$. We can go from binary vectors, that model interest/no interest for $Q$ and presence/no presence for $V_u$, to the more sophisticated case in which $Q$ expresses a weighted interest for some specific zones and $V_u$ is finely modeled by counting, for example, the number of events played by $u$ in each zone $z_i$. 
Now, given a query $Q$ and the players in the soccer-logs database, the goal is to design an algorithm that evaluates the \emph{propensity} of players to play in the field zones specified by $Q$. We follow the standard practice of Information Retrieval (IR) and compute for each player $u$ the dot product $s(u,Q) = V_u \cdot Q$. We can efficiently compute this product by means of one of the plethora of solutions known in the IR literature (see e.g. \cite{baeza,raghavan}). In this respect we point out that known solutions work efficiently over million (and more) dimensions, so that they {\em easily} scale to the problem size at hand, because $h\approx 10^6$ if we would assume zones $z_i$ of size $1\, cm^2$!

Finally {\sf PlayeRank} ranks players according to their \emph{rating} over a series of matches and their \emph{propensity} to play in the queried zones by sorting the players in decreasing order of the following score: 
\begin{equation}
    z(u, M, Q) = s(u, Q) \,\times\, \overline{r}(u, M)
\end{equation}
where $s(u, Q)$ is the dot product between $Q$ and the player vector $V_u$, and $\overline{r}(u, M)$ is $u$'s player rating over a series of matches. 
Note that the function $z(u, M, Q)$ could be defined in many other ways, for example by weighting $s(u,Q)$ and $\overline{r}(u, M)$ differently, in order to better capture the user's needs. Other combinations will be investigated in the future.

For the sake of presentation, we consider here a tessellation of the soccer field into 100 equal-sized zones and, thus, define a query $Q$ as a binary vector of $100$ components which express the interest of the user about the ``presence in a zone'' for the searched players. Then {\sf PlayeRank} computes $s(u, Q)$ as the dot product between $Q$ and the player vector $V_u$, and $\overline{r}(u, M)$ as the player rating over all matches of $u$. Then players are ranked in decreasing order of the quantity $z(u, M, Q) = s(u, Q) * \overline{r}(u, M)$ as described above.

\begin{figure}[htb]\centering
\includegraphics[scale=0.12]{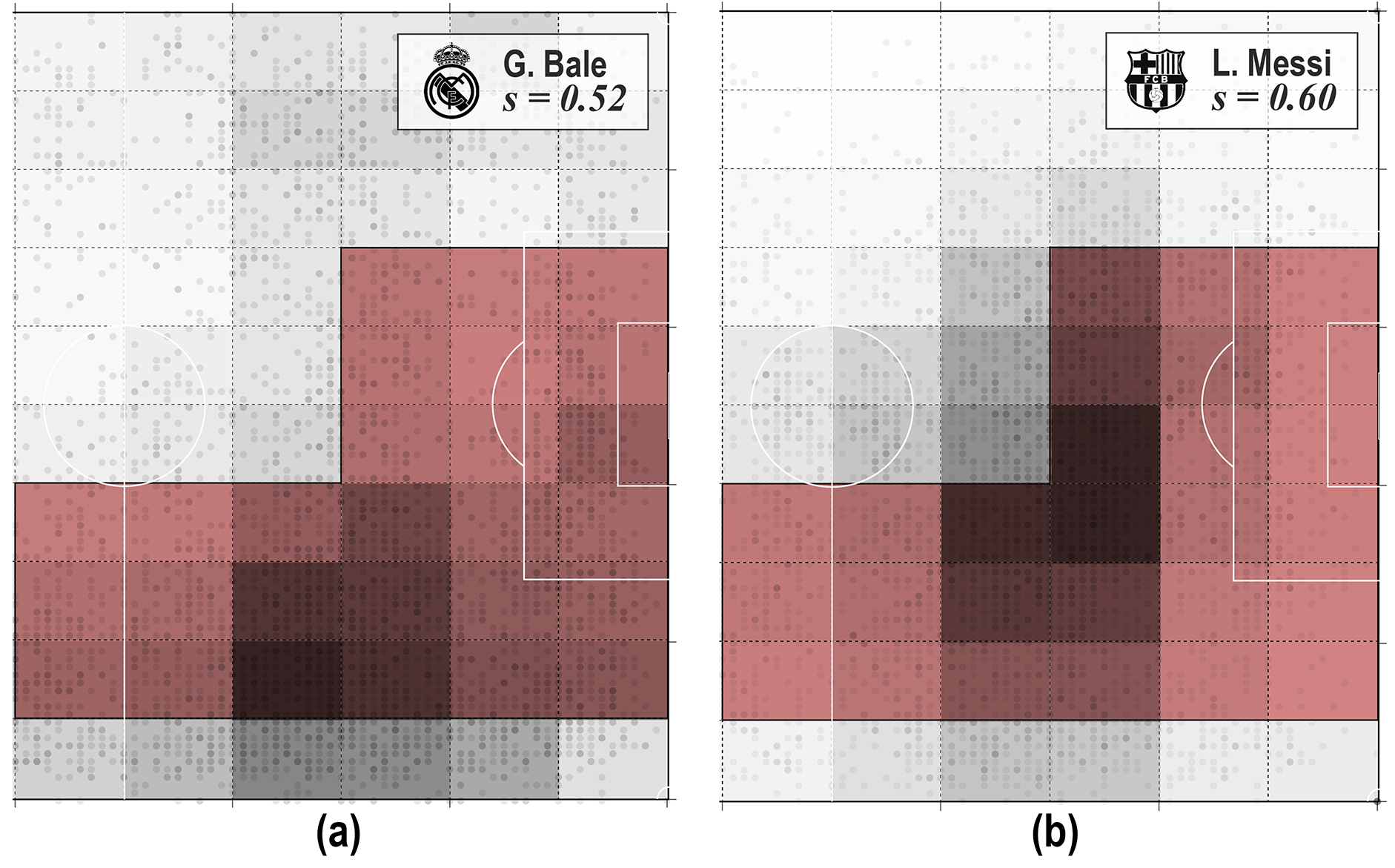}
\caption{Visualization of a spatial query $Q_1$ (red area) and the heatmaps of presence of Gareth Bale of Real Madrid (a) and Lionel Messi of Barcelona (b). The darker a zone the higher a player's propensity to play in it.}
\label{fig:query}
\end{figure}

Table \ref{tab:retrieve} shows the top-10 players in the our database according to their $z(u, M, Q_1)$ for an exemplar query $Q_1$ showed in Figure \ref{fig:query}. Lionel Messi, whose heatmap of positions is drawn in Figure \ref{fig:query}b, has the highest $z(u, M, Q_1)$. In the table, it is interesting to note that, though the vector of Arjen Robben is more similar to $Q_1$ ($s(Robben, Q_1) = 0.61$) than Messi's vector ($s(Messi, Q_1) = 0.60$), Messi has a higher player rating ($\overline{r}(Messi, M) = 0.46$, $\overline{r}(Robben, M) = 0.43$). As a result, the combination $z(u, M, Q_1)$ of the two quantities makes Messi the player offering the best trade-off between matching with the user-specified zones and performing well in those zones. 

\begin{table}[htb]\centering

\begin{tabular}{r | l || r | r | r || l}
   &               \bf player & $z$ &     $s$ & $\overline{r}$ & \bf club\\
   \hline
  1 &        L. Messi &  \bf 0.28 &  0.60 & \bf 0.46 & Barcelona\\
  2 &       A. Robben &  0.26 & \bf 0.61 &  0.43 & Bayern M.\\
  5 &   M. Salah &  0.24 &  0.56 &  0.43 & Liverpool\\
3 &  L. Su\'arez &  0.24 &  0.54 &  0.45 & Barcelona \\
4 &  T. M\"uller &  0.24 &  0.56 &  0.43 & Bayern M. \\
  6 &       R. Lukaku &  0.24 &  0.56 &  0.42 & Man. Utd\\
  7 &      A. Petagna &  0.23 &  0.55 &  0.42 & Atalanta\\
  8 &      D. Berardi &  0.22 &  0.54 &  0.41 & Sassuolo\\
  9 &          Aduriz &  0.22 &  0.55 &  0.40 & A. Bilbao\\
 10 &         G. Bale &  0.22 &  0.52 &  0.43 & R. Madrid\\
\hline
\end{tabular}
\caption{Top-10 players in the Wyscout DB according to their $z(u, M, Q_1)$ with respect to query $Q_1$ in Figure \ref{fig:query},  computed on the last four seasons of the five main European leagues (Serie A, La Liga, Bundesliga, and Premier League).}
\label{tab:retrieve}
\end{table}

\subsection{Versatility}
\label{sec:versatility}
The role detector of {\sf PlayeRank} enables the analysis of an important aspect of a player's behavior: his \emph{versatility}, that we define as a player's propensity to change role from match to match. To investigate this aspect, we define the versatility of a player as the Shannon entropy of his roles in a series of matches $M$: 
\begin{equation}
V(u, M) = - \frac{\sum_{i=1}^k p(u, M)_i \log p(u, M)_i}{\log k}
\end{equation}
where $k=8$ and $p(u, M)_i$ is the probability of player $u$ of playing in role $i$, computed as the ratio of the number of matches in $M$ in which $u$ played in role $i$. 

\begin{figure}
    \centering
    \includegraphics[scale=0.4]{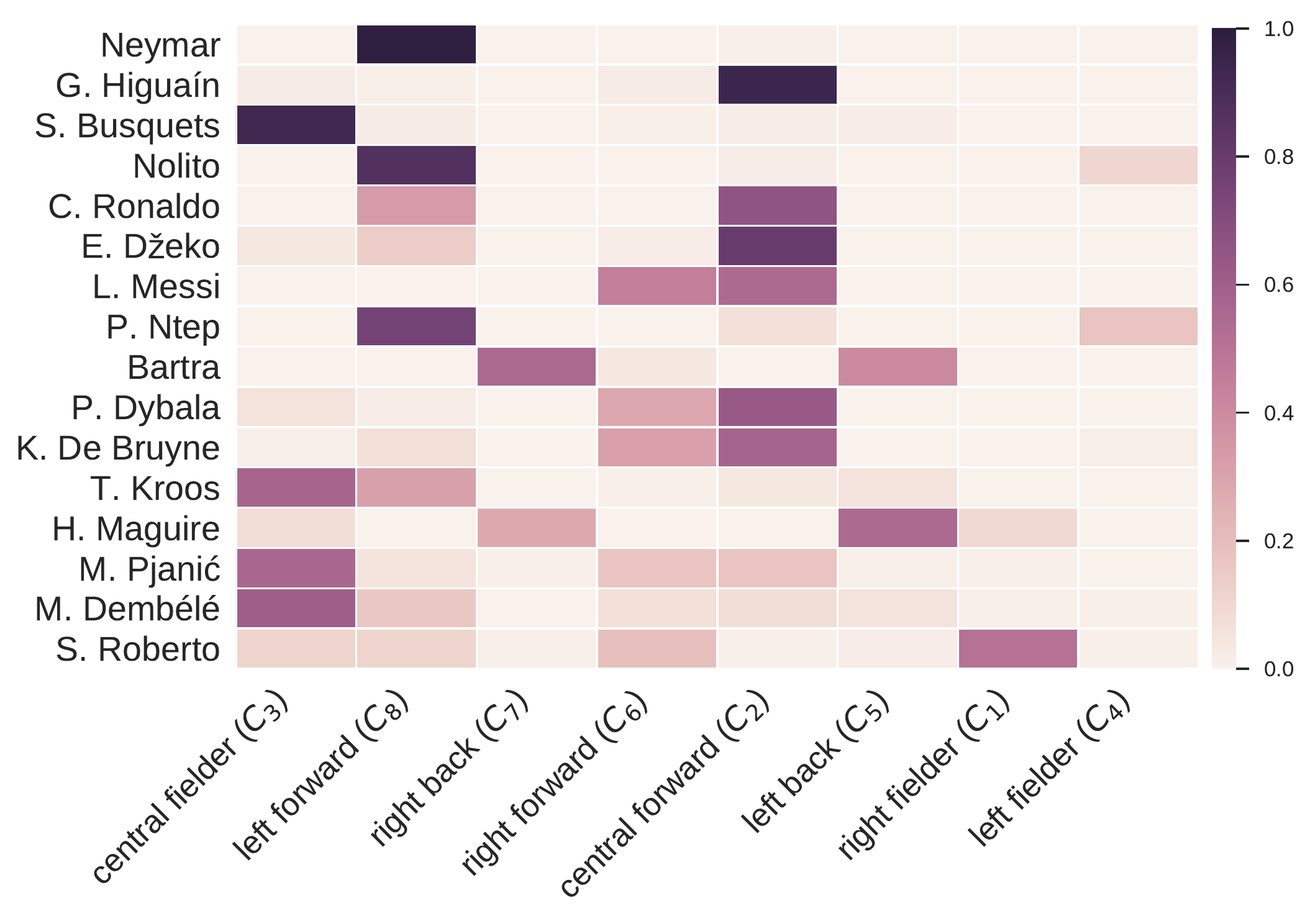}
    \caption{Heatmap showing the frequency of top players to play in the 8 roles (the darker a cell the higher the frequency). The players are sorted from the least versatile (Neymar) to the most versatile (Sergi Roberto).}
    \label{fig:versatility_heatmap}
\end{figure}

\begin{figure}
\centering
\includegraphics[scale=0.5]
{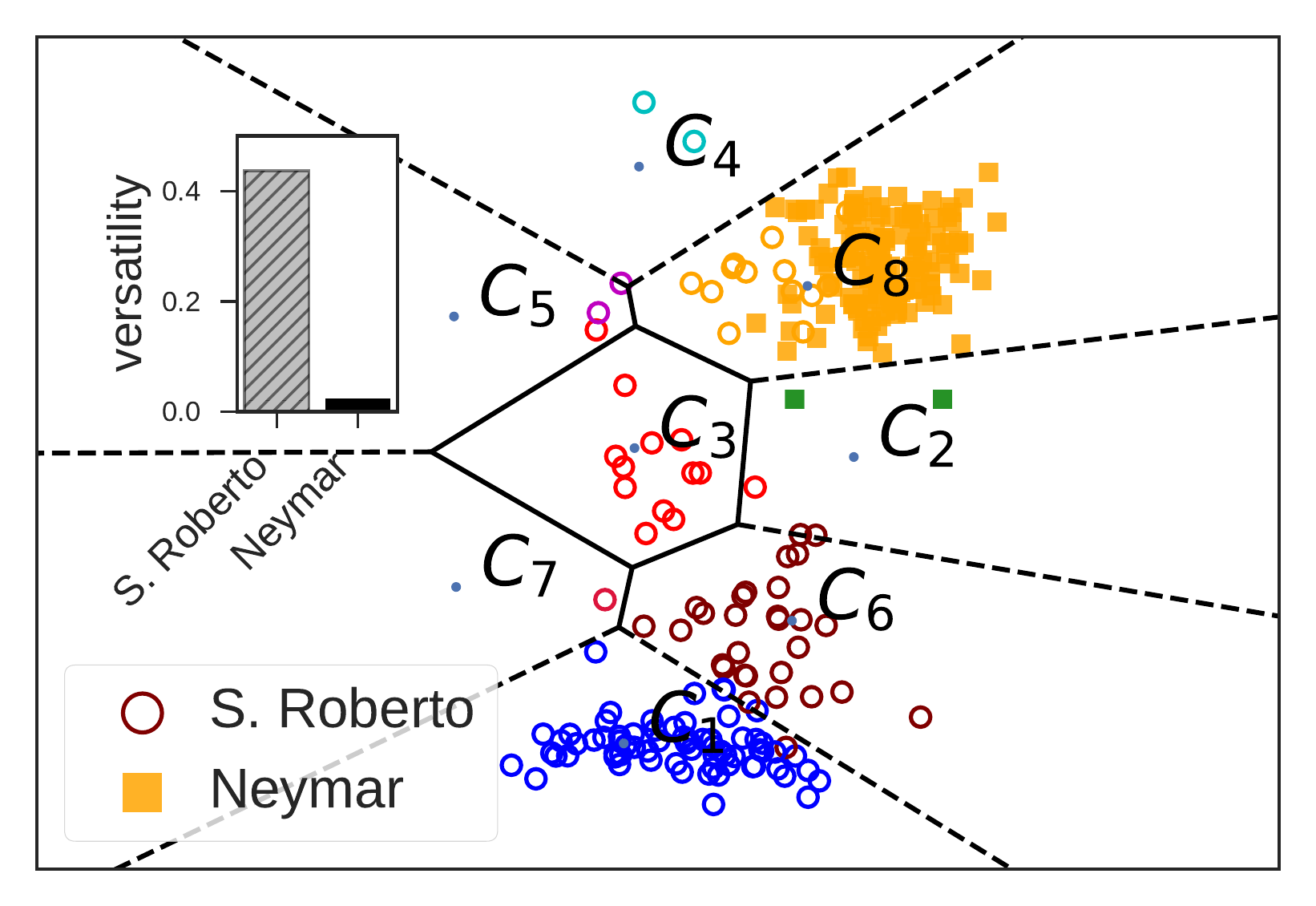}
\caption{ Positions of centers of performance of Sergi Roberto of Barcelona (circles) and Neymar of PSG (squares). Each center of performance is colored according to the role assigned by the role detection algorithm.}
\label{fig:versatility}
\end{figure}

Figure \ref{fig:versatility_heatmap} displays the frequency $p(u, M)_i$ of playing in a role $i$ for a set of top soccer players. We observe that many players have a high versatility, i.e., they play in different roles across different matches. In particular, Sergi Roberto (Barcelona) and Neymar (PSG) are among the most versatile and the least versatile players, respectively. Figure \ref{fig:versatility} visualizes all the centers of performance of Sergi Roberto and Neymar, coloring the centers according to the role assigned by the role detector. We observe that Neymar's centers of performance are concentrated in just one role ($C_8$, left forward) while Sergi Roberto's centers are scattered around the field, indicating that he plays in all 8 roles, witnessing a high versatility. Numerically, we observe that $V(\mbox{Sergi Roberto}) = 0.45$ and $V(\mbox{Neymar}) = 0.016$. The versatility of a player is an important property to take into account when composing a club's roster. {\sf PlayeRank} embeds versatility within its analytic framework, allowing soccer practitioners and scouters to evaluate the flexibility of a player as well as his playing quality in an automatic way.

\section{Conclusions and Future works}

In this paper we presented {\sf PlayeRank}, a data-driven framework that offers a multi-dimensional and role-aware evaluation of the performance of soccer players. Our extensive experimental evaluation on the massive database of soccer-logs provided by Wyscout -- 18 competitions, 31 million events, 21 thousands players -- showed that the rankings offered by {\sf PlayeRank} outperform existing approaches in being significantly more concordant with professional soccer scouts. Moreover, our experiments showed several interesting results, shedding light on novel patterns that characterize the performance of soccer players. 
Indeed we found that excellent performances are rare and unevenly distributed, since a few top players produce most of the observed excellent performances. An interesting result is also that top players do not always play excellence, they just achieve excellent performances more frequently than the other players. 
Regarding the extraction of feature weights, we found that the difference between the weights extracted from each competition separately is small (i.e., < 10\%) with the only exception of the Euro Cup and the World Cup for which that difference is slightly higher (i.e., $\approx 20\%$), thus highlighting the different nature of competitions for national teams. Lastly, our role detector found 8 main roles in soccer which we also exploited to investigate the versatility of players, an entropy-based measure which indicates the ability of a player to change role from match to match.

{\sf PlayeRank} is a valuable tool to support professional soccer scouts in evaluating, searching, ranking and recommending soccer players. 
We wish to highlight here that, given its modularity, {\sf PlayeRank} can be extended and customized in several ways. First, more sophisticated algorithms could be designed to detect a player's role during a match or fraction of a match. These algorithms could then be easily embedded in the {\sf PlayeRank}'s architecture, giving the user the possibility to customize role detection according to their needs. Some innovative AI-based solutions to role detection, that we plan to embed into {\sf PlayeRank}, have been recently proposed during a Soccer Data Challenge recently organized by Wyscout and SoBigData (\url{https://sobigdata-soccerchallenge.it/}). A similar reasoning applies to the feature weighting module: as soon as more sophisticated techniques will be proposed to weight performance features, they could be embedded in {\sf PlayeRank}'s architecture.

Another direction to improve {\sf PlayeRank} is to make it able to work with different data sources. In its current version, {\sf PlayeRank} is based on soccer-logs only, a standard data format describing all ball touches that occur during a match \cite{gudmundsson2017spatio, pappalardo2017quantifying}. Unfortunately, out-of-possession movements are not described in soccer-logs, making it difficult to assess important aspects  such as pressing \cite{andrienko2017visual} or the ability to create spaces \cite{fernandez2018wide}. {\sf PlayeRank} can be easily extended by making the individual performance extraction module able to extract features from other data sources like video tracking data \cite{gudmundsson2017spatio} and GPS data \cite{rossi2018effective}, which provide  a detailed description of the spatio-temporal trajectories generated by players during a match.

Finally, it would be interesting to investigate the flexibility of {\sf PlayeRank}'s architecture by plugging into it new performance metrics that will be proposed in the literature; as well as to evaluate its applicability to other team sports, such as basketball, hockey or rugby, for which data are available in the same format of soccer-logs \cite{gudmundsson2017spatio,stein2017how,rein2016bigdata}.

\begin{acks}
 This work is funded by EU project SoBigData RI, grant \#654024. 
We thank Daniele Fadda for support on data visualization and Alessio Rossi for his invaluable suggestions.
\end{acks}

\appendix
\section{Performance features}
Table \ref{tab:list_features} shows the list of features used in our experiments. Note that {\sf PlayeRank} is designed to work with any set of features, thus giving to the user a high flexibility about the description of performance. If other features are available from different data sources, describing for example physiological aspects of performance, they can be added into the framework. Section \ref{sec:validation} shows that the proposed set of features is powerful enough to make {\sf PlayeRank} outperform existing approaches in being more concordant with professional soccer scouts.

\begin{table*}[]
    \centering
    
    \begin{tabular}{l | l | l | l}
\hline
\bf type &                                   \bf  feature &   \bf       type &                                        \bf feature \\
\hline
\multirow{8}{*}{\rotatebox[origin=c]{90}{\textbf{duel}}}       &                     duel-air duel-accurate &  \multirow{11}{*}{\rotatebox[origin=c]{90}{\textbf{others on the ball}}}  &      others on the ball-accelleration-accurate \\
       &                 duel-air duel-not accurate &   &  others on the ball-accelleration-not accurate \\
       &        duel-ground attacking duel-accurate &   &          others on the ball-clearance-accurate \\
       &    duel-ground attacking duel-not accurate &   &      others on the ball-clearance-not accurate \\
       &        duel-ground defending duel-accurate &   &                others on the ball-touch-assist \\
       &    duel-ground defending duel-not accurate &   &        others on the ball-touch-counter attack \\
       &       duel-ground loose ball duel-accurate &  
       &   others on the ball-touch-dangerous ball lost \\
       &   duel-ground loose ball duel-not accurate &   &                 others on the ball-touch-feint \\
     \cline{1-2}
     
     \multirow{18}{*}{\rotatebox[origin=c]{90}{\textbf{foul}}}  &                    foul-hand foul-red card &  
     &          others on the ball-touch-interception \\
       &          foul-hand foul-second yellow card &  
       &           others on the ball-touch-missed ball \\
       &                 foul-hand foul-yellow card &   &           others on the ball-touch-opportunity \\
       \cline{3-4}
       &            foul-late card foul-yellow card &                \multirow{26}{*}{\rotatebox[origin=c]{90}{\textbf{pass}}}  &                       pass-cross pass-accurate \\
      &                  foul-normal foul-red card &                 &                         pass-cross pass-assist \\
       &        foul-normal foul-second yellow card &                 &                       pass-cross pass-key pass \\
       &               foul-normal foul-yellow card &                 &                   pass-cross pass-not accurate \\
      &             foul-out of game foul-red card &                 &                        pass-hand pass-accurate \\
       &   foul-out of game foul-second yellow card &                 &                    pass-hand pass-not accurate \\
       &          foul-out of game foul-yellow card &                &                        pass-head pass-accurate \\
       &                 foul-protest foul-red card &                 &                          pass-head pass-assist \\
       &       foul-protest foul-second yellow card &                &                        pass-head pass-key pass \\
       &              foul-protest foul-yellow card &                &                    pass-head pass-not accurate \\
       &    foul-simulation foul-second yellow card &                &                        pass-high pass-accurate \\
       &           foul-simulation foul-yellow card &                &                          pass-high pass-assist \\
       &                 foul-violent foul-red card &                &                        pass-high pass-key pass \\
       &       foul-violent foul-second yellow card &                &                    pass-high pass-not accurate \\
       &              foul-violent foul-yellow card &                &                      pass-launch pass-accurate \\
      \cline{1-2}
      
 \multirow{11}{*}{\rotatebox[origin=c]{90}{\textbf{free kick}}}  &        free kick-corner free kick-accurate &                 &                        pass-launch pass-assist \\
  &    free kick-corner free kick-not accurate &                 &                      pass-launch pass-key pass \\
  &         free kick-cross free kick-accurate &                 &                  pass-launch pass-not accurate \\
  &     free kick-cross free kick-not accurate &                 &                      pass-simple pass-accurate \\
  &        free kick-normal free kick-accurate &                &                        pass-simple pass-assist \\
  &    free kick-normal free kick-not accurate &                 &                      pass-simple pass-key pass \\
  &   free kick-penalty free kick-not accurate &                 &                  pass-simple pass-not accurate \\
  &          free kick-shot free kick-accurate &                 &                       pass-smart pass-accurate \\
  &      free kick-shot free kick-not accurate &                 &                         pass-smart pass-assist \\
  &      free kick-throw in free kick-accurate &                 &                       pass-smart pass-key pass \\
  &  free kick-throw in free kick-not accurate &                 &                   pass-smart pass-not accurate \\
  \cline{3-4}
        &                                         &                \multirow{2}{*}{\rotatebox[origin=c]{90}{\textbf{shot}}}  &                             shot-shot-accurate \\
        &                                         &                 &                         shot-shot-not accurate \\
\hline
\end{tabular}
    
    \caption{List of the 76 features extracted from the soccer-logs database and used in our experiments.}
    \label{tab:list_features}
\end{table*}

\bibliographystyle{ACM-Reference-Format}
\bibliography{biblio}

\end{document}